\documentclass[twocolumn]{aastex62}

\usepackage{amsmath}	
\usepackage{amssymb}	
\usepackage[table]{colortbl}
\usepackage{multirow, makecell} 

\newcommand{\ucell}{\cellcolor{red!75}}
\newcommand{\ocell}{\cellcolor{gray!75}}
\newcommand{\scell}{}

\usepackage[normalem]{ulem}

\newcommand{\f}[2]{\frac{#1}{#2}} 

\newcommand{\oldathena}{\textsc{Athena }}
\newcommand{\athena}{\textsc{Athena{\scriptsize ++ }}}
\newcommand{\athenapp}{\textsc{Athena{\scriptsize ++}}}

\newcommand{\beq}{\begin{equation}}
\newcommand{\seq}{\end{equation}}
\let\f=\ff
\newcommand{\f}[2]{\frac{#1}{#2}} 
\newcommand{\pd}[2]{\frac{\partial #1}{\partial #2}} 
\newcommand{\pdL}[1]{\left(\frac{\partial \mathcal{L}}{\partial T}\right)_{#1}} 
\newcommand{\pdAB}[2]{\left(\frac{\partial \Theta}{\partial #1}\right)_{#2}} 
\newcommand{\pdLrho}[1]{\left(\frac{\partial \mathcal{L}}{\partial \rho}\right)_{#1}}

\newcommand{\gv}[1]{\ensuremath{\mbox{\boldmath$ #1 $}}} 
\newcommand{\grad}[1]{\nabla #1} 
\renewcommand{\div}[1]{\nabla \cdot #1} 

\def \H{\gv{q}}

\def \kk{k^2}
\def \kkp{k^{'2}}
\def \kp{\gv{k}^{'}}
\def \Arho{A_\rho^{'}}

\def \Ap{A_p^{'}}

\def \Avv{\gv{A}_v^{'}}

\graphicspath{{./}{figures/}}

\received{October 25, 2018}
\submitjournal{ApJ}

\shorttitle{Non-isobaric TI}
\shortauthors{Waters \& Proga}

\begin{document}

\title{Non-isobaric thermal instability}

\correspondingauthor{Tim Waters}
\email{waters@lanl.gov}

\author{Tim Waters}
\affil{Theoretical Division, Los Alamos National Laboratory}

\author{Daniel Proga}
\affiliation{Department of Physics \& Astronomy, University of Nevada, Las Vegas} 

\begin{abstract}
Multiphase media have very complex structure and evolution. Accurate numerical simulations are necessary to make advances in our understanding of this rich physics. Because simulations can capture both the linear and nonlinear evolution of perturbations with a relatively wide range of sizes, it is important to thoroughly understand the stability of condensation and acoustic modes between the two extreme wavelength limits of isobaric and isochoric instability as identified by Field (1965). Partially motivated by a recent suggestion that large non-isobaric clouds can `shatter' into tiny cloudlets, we revisit the linear theory to survey all possible regimes of thermal instability. We uncover seven regimes in total, one of which allows three unstable condensation modes. Using the code \athenapp, we determine the numerical requirements to properly evolve small amplitude perturbations of the entropy mode into the nonlinear regime. Our 1D numerical simulations demonstrate that for a typical AGN cooling function, the nonlinear evolution of a single eigenmode in an isobarically unstable plasma involves increasingly larger amplitude oscillations in cloud size, temperature and density as the wavelength increases.  Such oscillations are the hallmark behavior of non-isobaric multiphase gas dynamics and may be observable as correlations between changes in brightness and the associated periodic redshifts and blueshifts in systems that can be spatially resolved. 
Intriguingly, we discuss regimes and derive characteristic cloud sizes for which the saturation process giving rise to these oscillations can be so energetic that the cloud may indeed break apart.  However, we dub this process `splattering' instead of `shattering', as it is a different fragmentation mechanism triggered when the cloud suddenly `lands' on the stable cold branch of the equilibrium curve.
\end{abstract}


\section{Introduction} \label{sec:intro}
The concept of thermal instability (TI) originated through an analysis of the heat equation by Parker (1953), who
envisioned a runaway scenario that would uniformly heat or cool the plasma.  The notion of condensations --- cold gas forming within a hotter background medium --- was introduced by Zanstra (1955), who emphasized the tendency for such two-phase media to maintain pressure equilibrium.  
The various instability criteria for TI were first derived by Field (1965) in a landmark paper.  

In the ensuing half century, it has become clear that condensation phenomena in very diverse astrophysical environments may be due to TI.   
For example, ALMA observations have provided strong evidence that TI operates in the central regions of cool-core clusters and within brightest cluster galaxies and brightest group galaxies (e.g., Russell et al. 2014; McNamara et al. 2014; David et al. 2014; Voit et al. 2015; Russell et al. 2016; Tremblay et al. 2016; Vantyghem et al. 2016; Temi et al. 2018; Pulido et al. 2018), as molecular gas must co-exist with the hot virialized plasma temperatures of the intracluster medium (ICM) or intergroup medium.
The relatively low temperature ($T \sim 10^4\,\rm{K}$) gas recently inferred to be present in the circumgalactic medium (CGM) of galaxies is likely also due to TI (Werk et al. 2013; Stocke et al. 2013; Stern et al. 2016; see Tumlinson et al. 2017 for a review). 
On smaller scales, TI is the core idea behind the earliest two- and three-phase models of the ISM (Field et al. 1969; McKee \& Ostriker 1977; for a review, see Cox 2005) and it is still a key element in star formation studies (e.g., Kim et al. 2013; Kim \& Ostriker 2015) and models of interstellar turbulence 
(e.g., Brandenburg \& Nordlund 2011; Choi \& Stone 2012; Iwasaki \& Inutsuka 2014; Kritsuk et al 2017).  
It has been shown that TI is the mechanism leading to multiphase accretion flows in simulations of generalized Bondi accretion that include heating and cooling processes (Barai et al. 2011, 2012); Gaspari et al. (2012) christened this mechanism `cold chaotic accretion' or CCA.  Clumpy winds, which can be considered the outflowing counterpart to CCA, may explain much of the phenomenology associated with active galactic nuclei (AGNs; e.g., Elvis 2017; see also Kurosawa \& Proga 2009).
The original application discussed by Parker (1953), namely solar prominences, continue to be modeled as condensation processes using, for example, 3D MHD simulations of TI (e.g., Xia et al. 2014; Xia \& Keppens 2016).

Considering how widespread the applications are, it is important to fully understand how TI operates.  Field himself discussed, both qualitatively and quantitatively, several extensions to the basic linear theory to account for the effects of rotation, expansion, density stratification, and anisotropic conduction in ideal MHD (Field 1965).  Subsequent analytic work has focused on extending the theory of TI to account for additional physics including buoyancy effects accompanying gravity in both dynamic flows (Balbus \& Soker 1989) and stratified plasmas (e.g., Defouw 1970; Balbus 1986; Malagoli et al. 1987; Binney et al. 2009); MHD effects in ideal (e.g., Hennebelle \& Passot 2006), non-ideal (Heyvaerts 1974;  Shadmehri et al. 2010) and partially ionized (Stiele et al. 2006; Fukue \& Kamaya 2007) plasmas; self-gravity (e.g., Gomez-Pelaez \& Moreno-Insertis 2002); and variable chemical composition (which leads to the chemo-thermal instability; Yoneyama 1972; Sabano \& Yoshi 1977).  Further extensions to TI are summarized by Nekrasov (2011).  

Early numerical studies were focused on understanding the nonlinear regime of basic TI, especially in the context of ISM turbulence (e.g., V{\'a}zquez-Semadeni et al. 2000; Kritsuk \& Norman 2002; Koyama \& Inutsuka 2004; Audit \& Hennebelle 2005; Piontek \& Ostriker 2004, 2005; Gazol et al. 2005; Brandenburg et al. 2007).  
Recently, numerical simulations have been performed to investigate the nonlinear effects of buoyancy in dynamic flows (e.g., Mo{\'s}cibrodzka \& Proga 2013) and stratified plasmas (e.g. McCourt et al. 2012; Sharma et al. 2012; Li \& Bryan 2014; Meece et al. 2015).  
Only a few studies have included radiation forces (Proga \& Waters 2015; Waters \& Proga 2016) or magnetic fields (e.g., Piontek \& Ostriker 2004, 2005; Sharma et al. 2010; Choi \& Stone 2012; Wagh et al. 2014; Ji et al. 2018), physics that will likely prove essential for building realistic models of the multiphase environments in AGNs and the CGM.  

In this paper, on the contrary, we simply focus on understanding a largely unexplored aspect of basic TI, 
namely the long-wavelength regime.  
In a plasma that is stable isochorically but unstable isobarically, only the entropy mode can condensate.  
Large thermally unstable perturbations have commonly been assumed to evolve in an isochoric fashion (e.g., Burkert \& Lin 2000),
irrespective of the stability criterion being satisfied.  
There is a need, therefore, to study the behavior of this entropy mode 
at ever larger wavelengths.
We perform hydrodynamical simulations to reveal the nonlinear dynamics of the entropy mode as it gradually changes from behaving isobarically to undergoing large changes in pressure and finally settling into a constant density core.
  
While it is known that isochoric instability is difficult to trigger for realistic astrophysical cooling functions (e.g., Balbus 1995), there are circumstances in which it can occur, for example when there is a deficit of soft X-rays in photoionized plasmas (see e.g., Figure~24 of Kallman \& McCray 1982).  Such a deficit shows up as regions of negative slope on the S-curves in the $[\log(T),\log(\xi)]$-plane, also found by Dyda et al. (2017; we define $\xi$ in \S{3}).  This occurrence is indicative of a doubly unstable regime in which the plasma satisfies both the isobaric and isochoric instability criteria (see \S{3.3}).  

It is of interest to examine the implications of forming clouds from large wavelength perturbations, in part to investigate 
the `shattering' mechanism proposed by McCourt et al. (2018; hereafter M+18).  By drawing an analogy with gravitational fragmentation, M+18 hypothesized that non-isobaric clouds will restore pressure balance by fragmenting into very small scale cloudlets.  
We demonstrate using 1D simulations that pressure equilibrium is instead regained by way of damped oscillations, even for very large wavelength perturbations, although we have identified a numerical effect that can be mistaken for `shattering' (see \S{5}).  Nevertheless, we do find that thermally unstable plasmas can be prone to a fragmentation process, albeit one quite different from `shattering'.  This new mechanism, which we call `splattering', arises because the saturation of TI is a sudden process in which the velocity of the condensating gas must reverse directions upon `landing' on the cold phase of the S-curve.  The velocity of gas feeding the condensation mode can, according to linear theory, exceed the sound speed at the end of the linear regime, so the cloud can fragment once the pressure of the reversed gas exceeds that of the confining gas.  

This paper is organized as follows.  In Section~2, we point out that there may be thermally unstable environments where isobaric clouds cannot even form in the first place.  In Section~3, we use linear theory to categorize the possible regimes of TI.  
In Section~4, we describe the results of 1D simulations designed to probe the nonlinear evolution of large wavelength entropy modes. 
In Section~5, we discuss the various regimes of TI uncovered, identifying which ones can lead to fragmentation, and we compare our 1D cloud sizes with the characteristic length scales appearing in the non-isobaric regime.  Finally, in Section~6 we summarize our results and note their observational implications.  In a companion paper, we present an application of non-isobaric gas dynamics, addressing how clouds of different sizes interact.  

\section{Basic considerations}
One of the reasons there has been so much focus on the isobaric regime of TI is that the growth rate of the entropy mode is inversely proportional to wavelength.   
That is, intuitively one would expect the fastest growing modes to be the most relevant in studies of multiphase gas dynamics.
However, 
it is yet to be determined if the dominance of fast growing modes lasts beyond the initial linear growth phase.  
Non-linear investigations are required to follow the evolution of multiphase gas with perturbations of various sizes to address questions such as `are the fastest forming clouds actually the ones that survive the longest?'. 

To quantify matters, consider the basic length scales of the problem.
Thermally unstable gas can condensate into clouds provided the perturbation
wavelengths $\lambda$ are sufficiently larger than
the Field length (Begelman \& McKee 1990), defined as
\beq \lambda_F \equiv 2\pi\sqrt{\f{\kappa T}{\rho \Lambda}}\bigg|_0, \label{lambda_F}\seq
where $\kappa$ is the thermal conductivity, $\rho$ and $T$ are the mass density and temperature of the plasma, and $\Lambda$ is the net cooling rate (defined in \S{3}).  The subscript `0' here and elsewhere denotes a quantity evaluated in the background equilibrium state, i.e. for \emph{the unstable plasma} out of which clouds condense.
Pre-existing clouds with characteristic dimensions $d_c$ less than $\lambda_F$ undergo classical evaporation (Cowie \& McKee 1977; Balbus 1985), while short wavelength perturbations ($\lambda \lesssim \lambda_F$) are stabilized by thermal conduction (Field 1965).  
Isobaric TI applies to perturbations on the order of a thermal length, $\lambda_{th}$, the distance sound waves propagate in a cooling time (e.g., Inoue \& Omukai 2015): 
\beq \lambda_{th} \equiv \left(c_s\,t_{cool}\right)_0, \label{lambda_th}\seq
where $c_{s} = \sqrt{\gamma k_BT/\bar{m}}$ (with $\bar{m}$ the mean particle mass) is the adiabatic sound speed and $t_{cool} \equiv \mathcal{E}/\Lambda$ is the timescale for gas to lose its thermal energy, $\mathcal{E} = c_{\rm{v}}\,T$.  

In certain environments such as AGNs, the isobaric regime is accompanied by cloud acceleration in the nonlinear phase of TI since the radiation field is highly anisotropic (Proga \& Waters 2015; hereafter, PW15).  This is due to the dramatic increases in bound-bound and bound-free opacity as the cloud forms.   
For our fiducial parameters in PW15, we found $\lambda_F= 0.19\, \lambda_{th}$.  The motivation for the present work
is the following consideration: in gas with $\lambda_{th} \lesssim \lambda_F$, perturbations that could form isobaric clouds are stabilized by conduction and therefore cannot condensate, implying that any clouds in such systems are non-isobaric.  Are the cooling rates that permit $\lambda_{th} < \lambda_F$ unrealistic?

A typical value for the cooling rate in both 
AGNs and the CGM is $L = 10^{-23}\,\rm{erg\,cm^3\,s^{-1}}$ (e.g., Kallman \& McCray 1982; Sutherland \& Dopita 1993).  
The cooling term $\rho \Lambda$ in equation \eqref{lambda_F} has units $\rm{erg\,cm^{-3}\,s^{-1}}$ and is related to $L$ as $\rho \Lambda = n_e n_H L$.\footnote{That is, $\Lambda = \mu (\mu_e \mu_H m_p)^{-1}  n L $, with $\mu m_p \equiv \rho/n = \bar{m}$, $\mu_H m_p \equiv \rho/n_H$, and $\mu_e m_p \equiv \rho/n_e$, where $n,n_e$ and $n_H$ denote the total particle, electron, and hydrogen number densities.  The symbol $n$ will hereafter denote perturbation growth/damping rates.} 
Expressing $\Lambda$ in terms of $L$ cancels the explicit density dependence of the ratio $\lambda_{th}/\lambda_F$, leaving only an implicit dependence through the function $L$ itself.  Further expressing the conductivity as $\kappa(T) = \chi T^{5/2}$, with $\chi =1.84\times 10^{-5}/\ln\Lambda_C~\rm{erg\,s^{-1}\,K^{-1}\,cm^{-1}}$ (the Spitzer value for a fully ionized plasma used by Cowie \& McKee 1977) gives,
\beq \f{\lambda_{th}}{\lambda_F} = \f{3.7}{\gamma - 1}  \left(\f{\ln\Lambda_C}{20}\f{\mu_e \mu_H}{\mu^3} \right)^{1/2} T_5^{-1/4} L_{23}^{-1/2} ,\label{thermal-Field-ratio}\seq
where $\ln\Lambda_C$ is the Coulomb logarithm, $T_5 = T/10^5\rm{K}$, and $L_{23} = L/10^{-23}~\rm{erg\,cm^3\,s^{-1}}$.
For a plasma of solar abundances, $\mu = 0.62$, $\mu_H = 1.43$, and $\mu_e = 1.18$, and we find $\lambda_{th}/\lambda_F = 14.8~T_5^{-1/4} L_{23}^{-1/2}$ when $\gamma = 5/3$ and $\ln\Lambda_C = 20$.  
Thus, for these values, the bound for isobaric clouds to not exist, $\lambda_{th} < \lambda_F$, is equivalent to 
\beq L \gtrsim 2\times 10^{-21}\,T_5^{-1/2} \,\rm{erg\,cm^3\,s^{-1}}.\seq
Such rates are not unrealistic in collisionally ionized environments with at least solar metallicities (see e.g., fig.~8 from Sutherland \& Dopita 1993).  However, in photoionized gas, this threshold is typically never reached unless number densities exceed $10^{12}\,\rm{cm}^{-3}$ (see e.g., fig.~2 of Nakayama \& Masai 2001; see also Gnedin \& Hollon 2012).  For example, Mehdipour et al. (2016), in a paper comparing three popular photoionization codes (\textsc{\small CLOUDY}, \textsc{\small XSTAR}, and \textsc{\small SPEX}) for various spectral energy distributions, found that the cooling rate never exceeds $L = 10^{-22}\,\rm{erg\,cm^3\,s^{-1}}$ for ionization parameters observed in AGN (see their fig.~5).

While the above calculation reveals that the Field length is always smaller than the wavelength of isobaric perturbations in photoionized environments, it has nonethless been found that isobaric clouds 
are highly prone to evaporation as they get disrupted.  This occurs due to, for example, velocity shear driven by radiation forces in the case of AGNs (PW15) or to wind-cloud interactions in galactic outflows (e.g., Br{\"u}ggen \& Scannapieco 2016).  Large clouds (those with $d_c >> \lambda_{th}$), meanwhile, can possibly survive much longer because cloud destruction timescales are typically proportional to the column density (Krolik et al. 1981).  

Additionally, we have found that the long term survival of isobaric condensations in an AGN environment is due to the continuous replenishment of multiphase gas, as the destruction of clouds ensures the presence of many perturbations that can re-condense (see Waters \& Proga 2016).  This mechanism requires the plasma to remain isobarically unstable at all times, while the conditions for TI are likely only triggered during active phases of AGN duty cycles (e.g., Gaspari et al. 2012).  Furthermore, they
may only be met at a certain range of radii for a limited period of time in dynamical environments (e.g., Mo{\'s}cibrodzka \& Proga 2013).  In this case, it may be necessary that large clouds can survive outside of the thermally unstable regions in order to satisfy basic mass budget constraints set by the strength of broad emission lines (e.g., Krolik 1999).  
Indeed, the CCA mechanism mentioned in \S{1} advocates such a picture (see also Voit et al. 2017), as more recently discussed by Gaspari \& S{\c a}dowski (2017), who base their conclusions on kiloparsec scale simulations of galactic halos demonstrating the presence of large, long-lived clouds (Gaspari et al. 2017; see also Gaspari 2015).  
Given these considerations, non-isobaric TI may turn out to be more relevant than initially suspected.

\begin{figure*}
\includegraphics[width=\textwidth]{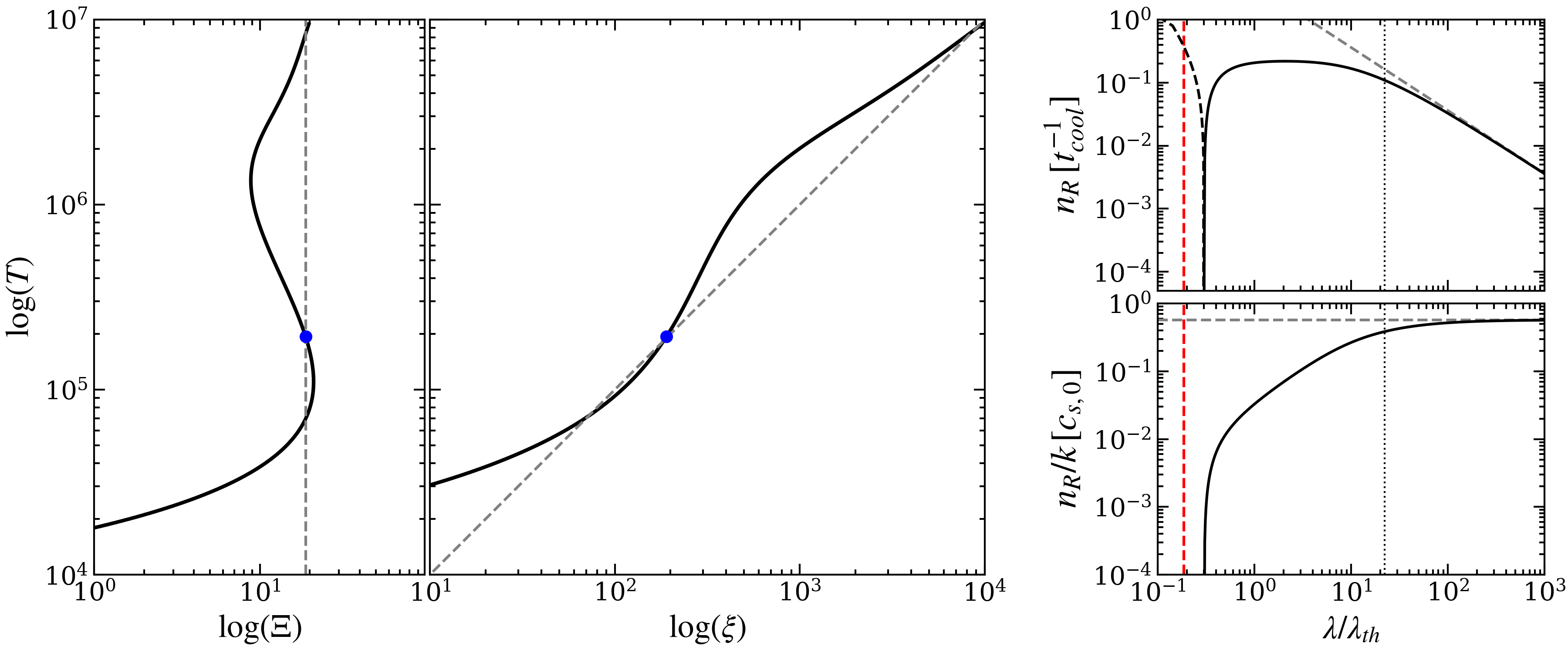}
\caption{The left two panels show the radiative equilibrium contours (S-curves) used in this work plotted against both the pressure and density ionization parameters, $\Xi$ and $\xi$, respectively.  The dashed gray lines indicate a constant pressure slope on either plot; isobaric perturbations will evolve toward the stable hot and cold branches that intersect these lines.   
All of our numerical results are calculated for the unstable locations marked with a blue dot.
The right two panels show $n_R$, the growth rate of the entropy mode, as well as $n_R/k$, as a function of perturbation wavelength $\lambda = 2\pi/k$.  
The dashed portion of $n_R$ denotes negative values, illustrating how TI is stabilized at short wavelengths comparable to the Field length (marked here with red lines).  Field (1965) showed that in the $k\rightarrow0$ (very large wavelength) limit of TI, $n_R/k$ approaches a constant for the entropy mode, corresponding to growth rates tending to 0.  The gray dashed lines in these plots are the asymptotic values of $n_R$ and $n_R/k$ as given by equation \eqref{n_max}; $R=-0.33$ at the blue dot.  The dotted vertical line marks $\lambda_c \approx 22.3 \,\lambda_{th}$, the wavelength at which acoustic modes transition into condensation modes.} 
\end{figure*}

\section{Linear Theory} 
While the results derived in this section are completely general, we adopt particular heating and cooling functions appropriate for high-temperature, photoionized environments such as AGNs to support our analysis.  By convention, an ionization parameter is often used in place of pressure or density when calculating phase diagrams for such heating and cooling rate functions.  For completeness, we define both the `density ionization parameter', $\xi \equiv 4\pi F_X \mu_H m_p/\rho$, where $F_X$ is the mean flux capable of ionizing hydrogen, and the `pressure ionization parameter', $\Xi \equiv (F_X/c)/p$, where $c$ is the speed of light and $p = \rho k_B T /\bar{m}$ is the gas pressure.  They are related as $\xi = 4\pi (\mu_H/\mu) c\, k_B T\,\Xi$ and can thus be used interchangeably.  Note that $\Xi$ is dimensionless, whereas $\xi$ has units of $\rm{erg\,cm\,s^{-1}}$.  

For a given $\xi$ and $T$, gas in radiative equilibrium --- when the cooling function $\Lambda = \Lambda(\xi,T)$ is balanced by a heating function, $\Gamma(\xi,T)$ --- defines an equilibrium contour on the $[\log(T),\log(\xi)]$-plane where $\mathcal{L} \equiv \Lambda - \Gamma = 0$.  This is commonly denoted the S-curve (e.g., Lepp et al. 1985).  
The top panels in Figure~1 show the net cooling function $\mathcal{L} = \mathcal{L}(\xi,T)$ used in this work, which was originally calculated by Blondin (1996).  Every point on these S-curves [indeed, on the entire $(T,\xi)$ or $(T,\Xi)$-plane] yields a different dispersion relation for TI, showing how the growth rate of the instability $n_R$ varies with wavenumber, $k = 2\pi/\lambda$.  The right panels in Figure~1 are plots of both $n_R$ and $n_R/k$ as a function of $\lambda/\lambda_{th}$ for one specific equilibrium location, marked by the blue dots on the S-curves.

\subsection{Isobaric and isochoric regimes}
The isochoric regime is commonly associated with the $k=0$ limit of TI,
but Field (1965) showed that there are two separate asymptotic limits.
Generalizing Field's cubic dispersion relation to account for perturbations off of the S-curve (see Appendix A), we arrive at 
\beq 
n\left(\f{n}{k}\right)^2 + \f{N_{\rho}}{t_{cool}}\left(\f{n}{k}\right)^2 + n\, c_{s,0}^2 + \f{N_p}{\gamma t_{cool}} c_{s,0}^2 = 0 ,
\label{DR}
\seq
where $N_{\rho}$ and $N_p$ are the dimensionless quantities
\beq
\begin{split}
N_{\rho} &\equiv \f{T_0}{\Lambda_0}\pdL{\rho} + \left(\f{\lambda_F}{\lambda}\right)^2, \\
N_p & \equiv \f{T_0}{\Lambda_0}\pdL{p} - \f{\mathcal{L}}{\Lambda_0} + \left(\f{\lambda_F}{\lambda}\right)^2 \\
        &= \f{T_0^2}{\Lambda_0} \left[\f{\partial (\mathcal{L}/T_0)}{\partial T}\right]_p + \left(\f{\lambda_F}{\lambda}\right)^2.
\end{split}
\label{NpNrho}
\seq
Here, the subscripts indicate which thermodynamic variable is held fixed when measuring net cooling changes in response to small increases in temperature.  To obtain this dispersion relation, which governs both the entropy mode and the two acoustic modes, we assumed perturbations of the form $\exp(n t + i\gv{k} \cdot \gv{x})$, where $n = n_R + in_I$ is complex valued.  We refer to a given mode as being stable if $n_R <0$, unstable if $n_R > 0$ and $n_I = 0$, and \emph{overstable} if $n_R > 0$ and $n_I \neq 0$. 

Notice from equation \eqref{DR} that the $k=0$ limit differs depending on whether or not $n$ simultaneously approaches 0.  
As shown by Field (1965), in this limit either $n$ or $n/k$ approaches a constant.
To see this clearly, let us focus only on the isobaric and isochoric condensation modes, which have $n_I = 0$.  
Then defining $u \equiv n_R/k$ we can formally rewrite equation \eqref{DR} as
\beq 
n_R = - \f{N_\rho}{t_{cool}} \f{u^2 + R\, c_{s,0}^2}{u^2 + c_{s,0}^2},
\label{n_general}
\seq 
where we have defined the ratio of the cooling rate derivatives, 
\beq R \equiv \f{N_p}{\gamma N_\rho} .\label{Rdef}\seq
It will be seen that critical values of this ratio distinguish the various regimes of TI.  

The isochoric $k=0$ limit is the one in which $n_R\rightarrow constant$ because the isochoric instability criterion follows from the $u\rightarrow \infty$ limit: $n_R > 0$ when $N_\rho < 0$, or (since $\lambda_F/\lambda$ vanishes in this limit)
\beq \pdL{\rho} < 0. \seq
In the limit where $u\rightarrow constant$ as $k\rightarrow 0$, the large-wavelength regime of the entropy mode is obtained instead. 
Further taking $n_R \rightarrow 0$ yields the values $u = \pm c_{s,0} \sqrt{-R}$, 
the positive value providing the asymptotic growth rate of the entropy mode\footnote{The negative value will be interpreted in \S{\ref{sec:lambda_crit}}.}
\beq n_R = k\,c_{s,0}\sqrt{-R}. \label{n_max}\seq
This growth rate was derived by Field (1965; see his equation 36).  However, he did not address the ambiguity with this expression, namely that the isochoric criterion also implies instability when $N_p > 0$ and $N_\rho < 0$.  We address this in \S{\ref{sec:lambda_crit}} but emphasize here that the isobaric instability criterion does in fact always determine the stability of the entropy mode.  This follows directly from equation \eqref{n_general}, but seeing how first requires insight into what $u$ represents physically.  

The analytic solutions to the equations of hydrodynamics describing the condensation process in the linear regime of TI 
are\footnote{Here, the velocity $v$ is defined as $\gv{v}\cdot\gv{k}/k$, where $k = |\gv{k}|$.} 
\beq
\begin{split}
\rho(\gv{x},t) &= \rho_0 + A \rho_0 e^{n_R t} \cos(\gv{k}\cdot{\gv{x}})  \\
v(\gv{x},t) &= - A\left(\f{n_R}{k}\right) e^{n_R t}\sin(\gv{k}\cdot{\gv{x}})  \\
p(\gv{x},t) &= p_0-A \rho_0 \left(\f{n_R}{k}\right)^2 e^{n_R t} \cos(\gv{k}\cdot{\gv{x}}),
\end{split}
\label{linear_profiles}
\seq
as derived in Appendix~A [see equation \eqref{linear_solutions}], where $A\equiv \delta\rho/\rho_0$ is a free parameter setting the initial density amplitude.  
These solutions apply in the comoving frame of a non-accelerating and initially homogenous region.  The meaning of the quantity $n_R/k$ becomes clear upon considering the end of the linear regime when the amplitude of the density perturbation $A\, e^{n_R t}$ becomes order unity.  From the velocity equation, we see that $n_R/k = u$ is the characteristic \emph{inflow speed} of gas feeding the condensation, while $\rho_0(n_R/k)^2$ is the pressure of the perturbation itself.  
In the isobaric regime, condensations grow by way of very small inflow velocities, $n_R/k << c_{s,0}$.  Thus, we can set $u = 0$ in equation \eqref{n_general} and arrive at the isobaric criterion for TI, $N_p < 0$ (by again demanding $n_R>0$).  By the definition of $N_p$, this is Balbus' criterion for TI (Balbus 1986), generalized to include the stabilizing effect of thermal conduction at sufficiently small $\lambda$:\footnote{This criterion was also obtained by Kim \& Narayan (2003) using a Lagrangian analysis.}
\beq  \left[\f{\partial (\mathcal{L}/T_0)}{\partial T}\right]_p < -\f{\Lambda_0}{T_0^2} \left(\f{\lambda_F}{\lambda}\right)^2.\label{FieldCriterion}\seq
Note that for $\lambda_F = 0$, this generalized isobaric criterion reduces to Field's original criterion for perturbations applied on the S-curve (where $\mathcal{L} = 0$).  It is important to include $\lambda_F$ in this regime, however, as the top right plot in Figure~1 shows.  The vertical red line marks the Field length, and $n_R$ changes sign in its vicinity, the dashed portion of $n_R$ representing the exponential decay rate.  

We can finally establish that the isobaric instability criterion also applies in the $k\rightarrow 0$ limit.  
Because the latter regime is entered gradually as $k$ decreases, the quantity $u^2$ in equation \eqref{n_general} is always less than $|R|\,c_{s,0}^2$ and only approaches it asymptotically.  Thus, $n_R$ is always positive provided $R < 0$ (and $N_\rho > 0$), so the isobaric instability criterion, as given in equation~\eqref{FieldCriterion}, still governs the entropy mode at very large wavelengths.

Let us now consider what the analytic solution in equation \eqref{linear_profiles} reveals about the transition from isobaric to non-isobaric TI.  
In the isobaric regime, we must have $u << c_{s,0}$ to recover the known growth rate of the entropy mode, $n_R = -(N_p/\gamma) t_{cool}^{-1}$, from equation \eqref{n_general}.  For consistency, therefore, it must be that $|N_p/\gamma|$ is significantly less than 1 for $k \sim \lambda_{th}^{-1}$ per our earlier discussion relating to equation \eqref{lambda_th}. 
This is indeed the case for our cooling function, as can be seen by comparing the plots on the right in Figure~1.  Notice that the largest TI growth rates have $n_R/k < 0.1 c_{s,0}$, and that the conduction term in equation \eqref{NpNrho} is responsible for the steep falloff in $n_R/k$ at smaller $\lambda$.  As the non-isobaric regime is gradually entered, the inflow velocity increases, but $u/c_{s,0}$ is never to exceed $\sqrt{-R}$, the maximum Mach number that the condensating gas can have by equation \eqref{n_max}.  
At the blue dot on our S-curve, we find $n_R/k \rightarrow 0.58\, c_{s,0}$, marked by the dashed horizontal line in the plot of $n_R/k$ in Figure~1.

Lastly, consider isochoric TI, the regime in which $n_R\rightarrow constant$ as $k\rightarrow 0$.  Equation \eqref{linear_profiles} shows that the velocity and pressure both diverge as $k^{-1}$, so any linear regime evolution must have an amplitude small enough that $A\, e^{n_R t}(n_R/k)$ is subsonic; otherwise the evolution will be fully nonlinear.  Moreover, the solution for $\rho(\gv{x},t)$ shows that the density cannot possibly remain constant unless the linear regime of exponential density growth is very short lived.  As Field (1965) commented, isochoric evolution is inconsistent with the force equation.  
We further point out that cloud formation in isochorically unstable plasmas
can proceed with condensation velocities exceeding the local sound speed,
and this nonlinear effect may lead to fragmentation (see \S{5}).

\subsection{The fast and slow isochoric condensation modes}\label{sec:lambda_crit}
Compressional, \emph{adiabadic} hydrodynamics always has two acoustic modes and one stationary entropy mode.  The existence of two (non-propagating) condensation modes with different asymptotic properties, namely $n_R\rightarrow constant$ vs. $n_R\propto k$ as $k\rightarrow 0$, implies that the large-wavelength regime of TI is accompanied by a rather extreme phenomenon: the loss of large-wavelength sound waves.  We will see below that this occurs beyond a critical wavelength whenever the plasma is unstable by the isobaric or isochoric criterions alone, while it can only occur over a limited range of wavelengths when the plasma is unstable by both criterions. 

Recall that cubic dispersion relations alternatively admit three real-valued solutions, hence the acoustic modes must turn into two additional condensation modes at some critical wavelength, denoted $\lambda_c$.  
To derive $\lambda_c$, note that equation \eqref{DR} transitions from having one real and one complex-valued pair of roots to all real roots when (see e.g., Press et al. 1999) 
\beq
\left[ \f{N_\rho^2}{9} - \f{1}{3}k_c^2\lambda_{th}^2 \right]^3 
=  \left[ \f{N_\rho}{2} k_c^2\lambda_{th}^2 \left(R - 1\right) +  \f{N_\rho^3}{27} \right]^2.
\label{R2equalsQ3}
\seq
This leads to an equation quadratic in $k_c^2$, giving the critical wavelengths 
\beq \lambda_\pm^2 = \f{8\pi^2}{N_\rho^2} \f{\lambda_{th}^2}{B}\left[-1 \pm \sqrt{1 - \f{4R}{B^2}} \right]^{-1},  \label{lambda_pm}\seq
where 
\beq B = \f{27}{4}\left(R - \f{1}{3}\right)^2 - 1.\seq
Only one of the roots $\lambda_\pm$ is real valued when $R<0$, while both are real when $0 < R < 1/9$; positive $R$ cases are examined in \S{3.3}.  The critical wavelength $\lambda_c$ is therefore uniquely defined as the real valued root when $R < 0$.  Since this occurs at small $k_c$, to a good approximation we can drop the $k_c^4$ term and arrive at\footnote{This approximation also follows from equation \eqref{lambda_pm} (for the plus sign) when $4R/B^2 << 1$ and it breaks down before reaching $B=0$, as $k$ is no longer small.  The precise values are $\lambda_c = \lambda_+$ for $R< 1/3 - 2/(3\sqrt{3}) \approx -0.05$ and $\lambda_c = \lambda_-$ for $-0.05 \lesssim R \leq 0$.}
\beq  \lambda_c \approx \f{2\pi}{|N_\rho|}\lambda_{th} \sqrt{-B/R}. \label{lambda_crit} \seq
The square root term is seen to be order unity everywhere along our S-curve. 
The vertical dotted lines in Figure~1 mark $\lambda_c$, which evaluates to $22.3\,\lambda_{th}$ at the blue dot on our S-curve (where $N_\rho = 0.69$ and $R=-0.33$).

\begin{table*}[]
\centering
\label{Table}
\begin{tabular}{c||*{12}{c|}c|c} 
\hline
\hline
Rate dependence & \multicolumn{6}{c|}{Wavelength dependence} & \multicolumn{6}{c|}{Mode Type \& Stability}  & Mode & $n_R\,t_{cool}$ ($\lambda \rightarrow \infty$)  \\ 
\hline
\hline
\multirow{3}{*}{\parbox{2cm}{\centering $R<0$ $(N_p < 0)$ $(N_\rho > 0)$}} 
&\multicolumn{3}{c|}{}   & \multicolumn{3}{c|}{} 
&\multicolumn{3}{c|}{\ucell condensation}   & \multicolumn{3}{c|}{\ucell condensation} & E & $k'\sqrt{-R}$ \\ 
&\multicolumn{3}{c|}{$\lambda < \lambda_c \text{                }$}   & \multicolumn{3}{c|}{$\lambda > \lambda_c$} 
&\multicolumn{3}{c|}{\scell acoustic}   & \multicolumn{3}{c|}{\scell condensation} & S & $-k'\sqrt{-R}$ \\ 
&\multicolumn{3}{c|}{}   & \multicolumn{3}{c|}{}
&\multicolumn{3}{c|}{\scell acoustic}   & \multicolumn{3}{c|}{\scell condensation} & F & $-N_\rho$ \\ \cline{1-15}
\multirow{3}{*}{\parbox{2cm}{\centering $R<0$ $(N_p > 0)$ $(N_\rho < 0)$}} 
&\multicolumn{3}{c|}{}   & \multicolumn{3}{c|}{}
&\multicolumn{3}{c|}{\scell condensation}   & \multicolumn{3}{c|}{\scell condensation} & E & $-k'\sqrt{-R}$ \\ 
&\multicolumn{3}{c|}{$\lambda < \lambda_c$}   & \multicolumn{3}{c|}{$\lambda >\lambda_c$}
&\multicolumn{3}{c|}{\ocell acoustic}   & \multicolumn{3}{c|}{\ucell condensation} & S & $k'\sqrt{-R}$  \\
&\multicolumn{3}{c|}{}   & \multicolumn{3}{c|}{}
&\multicolumn{3}{c|}{\ocell acoustic}   & \multicolumn{3}{c|}{\ucell condensation} & F & $-N_\rho$   \\
\hline
\hline
\multirow{3}{*}{\parbox{2cm}{\centering $0< R < \f{1}{9}$ $(N_p,N_\rho < 0)$}} 
&\multicolumn{2}{c|}{}  &\multicolumn{2}{c|}{}  & \multicolumn{2}{c|}{} 
&\multicolumn{2}{c|}{\ucell condensation}  &\multicolumn{2}{c|}{\ucell condensation}  & \multicolumn{2}{c|}{\ocell acoustic} & E & $\kkp(R - 1)/(2N_\rho)$  \\
&\multicolumn{2}{c|}{$\lambda < \lambda_-$}  &\multicolumn{2}{c|}{$\lambda_- < \lambda < \lambda_+$}  & \multicolumn{2}{c|}{$\lambda > \lambda_+$} 
&\multicolumn{2}{c|}{\ocell acoustic}  &\multicolumn{2}{c|}{\ucell condensation}  & \multicolumn{2}{c|}{\ocell acoustic} & S & $\kkp (R - 1)/(2N_\rho)$  \\
&\multicolumn{2}{c|}{}   &\multicolumn{2}{c|}{}   & \multicolumn{2}{c|}{} 
&\multicolumn{2}{c|}{\ocell acoustic}   &\multicolumn{2}{c|}{\ucell condensation}   & \multicolumn{2}{c|}{\ucell condensation} & F &  $-N_\rho$ \\ \cline{1-15}
\multirow{3}{*}{\parbox{2cm}{\centering $\f{1}{9} \leq R < \f{1}{3}$ $(N_p,N_\rho < 0)$}} 
&\multicolumn{3}{c|}{}   & \multicolumn{3}{c|}{}
&\multicolumn{3}{c|}{\ucell condensation}   & \multicolumn{3}{c|}{\ocell acoustic} & E & $\kkp (R - 1)/(2N_\rho)$  \\
&\multicolumn{3}{c|}{$\lambda < \lambda_o \text{                }$}   & \multicolumn{3}{c|}{$\lambda > \lambda_o$} 
&\multicolumn{3}{c|}{\ocell acoustic}   & \multicolumn{3}{c|}{\ocell acoustic} & S & $\kkp (R - 1)/(2N_\rho)$  \\
&\multicolumn{3}{c|}{}   & \multicolumn{3}{c|}{} 
&\multicolumn{3}{c|}{\ocell acoustic}   & \multicolumn{3}{c|}{\ucell condensation} & F & $-N_\rho$  \\  \cline{1-15}
\multirow{3}{*}{\parbox{2cm}{\centering $\f{1}{3}\leq R<1$ $(N_p,N_\rho < 0)$}} 
& \multicolumn{6}{c|}{} 
& \multicolumn{6}{c|}{\ocell acoustic} & E & $\kkp (R - 1)/(2N_\rho)$   \\
& \multicolumn{6}{c|}{None} 
& \multicolumn{6}{c|}{\ocell acoustic} & S & $\kkp(R - 1)/(2N_\rho)$  \\
&\multicolumn{6}{c|}{} 
&\multicolumn{6}{c|}{\ucell condensation} & F & $-N_\rho$\\  \cline{1-15} 
\multirow{3}{*}{\parbox{2cm}{\centering $R\geq1$ \\$(N_p,N_\rho < 0)$}} 
& \multicolumn{6}{c|}{} 
& \multicolumn{6}{c|}{\scell acoustic} & E & $\kkp (R - 1)/(2N_\rho)$  \\
& \multicolumn{6}{c|}{None}
& \multicolumn{6}{c|}{\scell acoustic} & S & $\kkp (R - 1)/(2N_\rho)$   \\
&\multicolumn{6}{c|}{}
&\multicolumn{6}{c|}{\ucell condensation} & F & $-N_\rho$ \\  
\hline
\hline
\multirow{3}{*}{\parbox{2cm}{\centering $R>1$ \\$(N_p,N_\rho > 0)$}} 
& \multicolumn{6}{c|}{} 
& \multicolumn{6}{c|}{\ocell acoustic} & E & $\kkp(R - 1)/(2N_\rho)$  \\
& \multicolumn{6}{c|}{None}
& \multicolumn{6}{c|}{\ocell acoustic} & S & $\kkp (R - 1)/(2N_\rho)$   \\
&\multicolumn{6}{c|}{}
&\multicolumn{6}{c|}{\scell condensation} & F & $-N_\rho$ \\  
\hline
\hline
\end{tabular}
\caption{Summary of the mode behavior and asymptotic growth/damping rates for classical TI.  The labels `E', `S', and `F' denote the entropy mode, the slow isochoric mode, and the fast isochoric mode, respectively (see \S{3.2}).  Stability depends on both the value of $R \equiv N_p/\gamma N_\rho$ and on a mode's wavelength when $R<1/3$ ($\lambda > \lambda_F$ is assumed).  The critical wavelengths $\lambda_c$, $\lambda_-$, $\lambda_+$, and $\lambda_o$ are defined in equations \eqref{lambda_crit}, \eqref{lambda_pm}, and \eqref{lambda_o}, and $k' \equiv k\lambda_{th} = 2\pi \lambda_{th}/\lambda$.  For any given $(R,\lambda)$, modes that are stable have no color shading, overstable acoustic modes (with $n_R > 0$ and $n_I \neq 0$) are shaded gray, and unstable condensation modes (with $n_R > 0$ and $n_I = 0$) are shaded red.  The asymptotic value of $n_R$ depends only on whether $R < 0$ or $R > 0$.} 
\end{table*}

The two modes that are acoustic modes for $\lambda < \lambda_c$ are condensation modes subject to instability for $\lambda > \lambda_c$.  They are both stable when $N_p < 0$ and both unstable when $N_\rho < 0$, i.e. they obey the isochoric instability criterion.   
To see this, notice that equation \eqref{n_max} implies that $n_R$ is positive whenever $R < 0$, regardless of whether $R < 0$ because $N_p < 0$ or $N_\rho < 0$.  When $R < 0$ due to isochoric instability ($N_\rho < 0$), the growth rate of the entropy mode is negative by definition.  It must be, therefore, that equation \eqref{n_max} gives the growth rate of one of the other two condensation modes.  
Recall from equation \eqref{n_general} that there is a second solution in the $n_R \rightarrow 0$, $k \rightarrow 0$ limit, namely
\beq n_R = -k\,c_{s,0}\sqrt{-R}, \label{n_inf}\seq
and this is the damping rate of the stable entropy mode.  
The remaining asymptotic growth rate is $n_R = -N_\rho/t_{cool}$, the growth rate associated with the isochoric condensation mode as identified by Field (1965).  Since $n_R/k$ is a measure of condensation velocity (see \S{2.1}), it seems appropriate to label the mode with $n_R/k = -k^{-1}N_\rho/t_{cool}$ the fast isochoric mode and the other (with $n_R/k = constant$) the slow isochoric mode.  By the same argument, these modes are stable when $R < 0$ due to the isobaric criterion ($N_p < 0$), with damping rates given by $n_R = -N_\rho/t_{cool}$ and equation \eqref{n_inf}.

\subsection{Combined instability: $R > 0$ cases}\label{sec:posR}
When $R > 0$ the plasma is either thermally stable or unstable by \emph{both} the isochoric and isobaric criterions.
In \S{\ref{sec:lambda_crit}} we showed that when $R < 0$, $n_I = 0$ for all of the modes in a plasma once $\lambda > \lambda_c$.  When $R > 0$, $n_I$ is always nonzero for two of the modes provided $R>1/9$, where $1/9$ is the solution to $4R = B^2$, the condition met when $\lambda_+ = \lambda_-$ [see equation \eqref{lambda_pm}].
For $R < 1/9$, the behavior of plasma modes now depends on both $\lambda_+$ and $\lambda_-$.  
Namely, for the small parameter space $0 < R < 1/9$, there again exist three condensation modes for $\lambda_- < \lambda < \lambda_+$ --- and they are all unstable.  For $\lambda < \lambda_-$, however, both the fast and slow isochoric modes are overstable with $n_I \neq 0$, the entropy mode being unstable with $n_I = 0$.  For $\lambda > \lambda_+$, on the other hand, both the entropy mode and slow isochoric modes are overstable, while the fast isochoric mode is unstable.   
 
For $1/9 < R < 1/3$ (where the critical value 1/3 follows momentarily), $\lambda_+$ and $\lambda_-$ no longer exist and a different characteristic wavelength, $\lambda_o$, distinguishes mode behavior.   
By the pattern discussed above, for $\lambda < \lambda_o$ we would expect both the fast and slow isochoric modes to be overstable, while for $\lambda > \lambda_o$,  the entropy and slow isochoric modes should be overstable instead.  This can indeed be verified numerically by directly solving the cubic dispersion relation, and therefore $\lambda_o$ marks where $n_I = 0$ for both the entropy mode and the fast isochoric mode.  Equating the real parts of the analytic solutions for cubic equations gives 
\beq \lambda_o = \f{6\pi}{|N_\rho|} \lambda_{th}\sqrt{\f{3}{2}\left(\f{1}{3} - R\right)}, \label{lambda_o} \seq
revealing the change in mode behavior for $R \geq 1/3$.  

The final critical $R$ value is $1$.  For $1/3 \leq R < 1$, the real valued mode is always the fast isochoric mode, with the slow isochoric mode and the entropy mode both being overstable.  For $R = 1$, corresponding to $N_p = \gamma N_\rho$, these two acoustic modes both become adiabadic with $n_R = 0$.  
For $R > 1$, the fast isochoric mode remains the only unstable mode; the two acoustic modes are now damped according to $n_R < 0$.  
Table~1 summarizes this ($R$,$\lambda$) dependence, providing an overview of all of the possible regimes of TI.  

We now determine the asymptotic growth rates of these modes to reveal the critical value $R=1$.  In Appendix~A, we derive the counterpart to equation \eqref{n_general} when $n_I \neq 0$:
\beq n_R = \f{N_\rho}{t_{cool}} \f{ (R/2) c_{s,0}^2 -\left[u^2 + (N_\rho/t_{cool}) u/k + c_{s,0}^2/2\right]}{4u^2 + 3(N_\rho/t_{cool}) u/k + c_{s,0}^2}. \label{nR_Rpos}\seq
We see that the limit $u\rightarrow constant$ as $k\rightarrow 0$ no longer describes the entropy mode.  Rather, $u/k\rightarrow constant$ does, implying that the $u^2$ terms are negligible in this limit, giving
\beq n_R \rightarrow \f{N_\rho}{t_{cool}} \f{ (R/2) c_{s,0}^2 -\left[(N_\rho/t_{cool}) u/k + c_{s,0}^2/2\right]}{3(N_\rho/t_{cool}) u/k + c_{s,0}^2}. \seq
Further requiring that $n_R \rightarrow 0$ in the $k=0$ limit gives
\beq \f{u}{k} N_\rho \rightarrow \f{t_{cool}}{2}c_{s,0}^2\left(R - 1\right) . \seq 
As asserted above, overstability of the entropy mode requires $0 < R < 1$ (because $N_\rho$ is negative by assumption).
Its asymptotic growth rate is
\beq n_R =\f{t_{cool}}{2 N_\rho}c_{s,0}^2\left(R - 1\right)  k^{2} . \label{Rpos_limit}\seq 
Notice that $n_R \propto k^2$, whereas for $R < 0$, the entropy mode has $n_R \propto k$ instead.  
The asymptotic growth rate of the slow isochoric mode is also given by \eqref{Rpos_limit}.

The fast isochoric mode is always real valued in the $k\rightarrow 0$ limit for all $R$, and so its asymptotic growth rate is always $n_R \rightarrow - N_\rho/t_{cool}$ by equation \eqref{n_general}.  However, equation \eqref{nR_Rpos} governs its growth rate in the instances in which it is overstable (see Table~1).  

\subsection{Isentropic instability}
We have made no mention of the isentropic criterion, which governs the stability of sound waves, because the critical wavelengths $\lambda_\pm$ and $\lambda_o$ already determine if there are modes with $n_I \neq 0$.  Our analysis has therefore fully accounted for all the ways that acoustic modes can become overstable, except for one: when the gas is both isobarically and isochorically stable (i.e., when $N_p > 0$ and $N_\rho > 0$), it can still be isentropically unstable.  
This situation is the bottom entry in Table~1, which is derived presently. 
Of course, all of the other instances of overstable acoustic modes in Table~1 must also be consistent with the isentropic criterion.  
This criterion is
\beq \pdL{s} < 0 \seq
(see Field 1965), where $s$ denotes entropy.  Using standard thermodynamic relations (e.g., see the appendix of Balbus 1995), we can express this criterion in terms of $N_p$ and $N_\rho$ as
\beq \gamma N_\rho (1-R) < \left(\f{\lambda_F}{\lambda}\right)^2 + \f{\mathcal{L}}{\Lambda_0} .\seq
If we focus only on the stability of wavelengths with $\lambda >> \lambda_F$ and for points on the S-curve (where $\mathcal{L} = 0$), this reduces to
\beq N_\rho (1-R) < 0 .\seq
We see that the stability of acoustic modes depends on whether $R < 1$ or $R > 1$, as well as on the sign of $N_\rho$.  For $R < 1$, acoustic modes are overstable whenever $N_\rho < 0$, i.e. whenever there is isochoric instability.  This covers all the cases in Table~1 from the second through the third to last rows.\footnote{This assumes that acoustic modes exist and is therefore also consistent with the 3rd row of Table~1 when $\lambda_+ < \lambda < \lambda_-$.}
When $R > 1$, this inequality reads $N_\rho > 0$, which corresponds to the cases in the bottom two rows of Table~1.  

The takeaway point in regards to acoustic modes is that in the presence of TI, the entropy mode is always acoustic at some wavelengths when $R > 0$.  In contrast, in adiabadic hydrodynamics, the entropy mode is always non-propagating and is therefore distinct from sound waves.  
It was necessary to introduce the separate classifications for `fast' and `slow' isochoric modes in order to track which modes are acoustic at any value of $(R,\lambda)$.  
It is helpful to view non-adiabadic hydrodynamics not as a system always consisting of two acoustic modes and an entropy mode that is advected with the flow, but rather as a potentially thermally unstable system that always consists of three modes prone to condensation.  
The acoustic modes need not always exist and when they do exist, they \emph{never} stay assigned to only the fast and slow isochoric modes at all wavelengths for any given value of $R$, as shown in Table~1.

\begin{figure}
\includegraphics[width=0.5\textwidth]{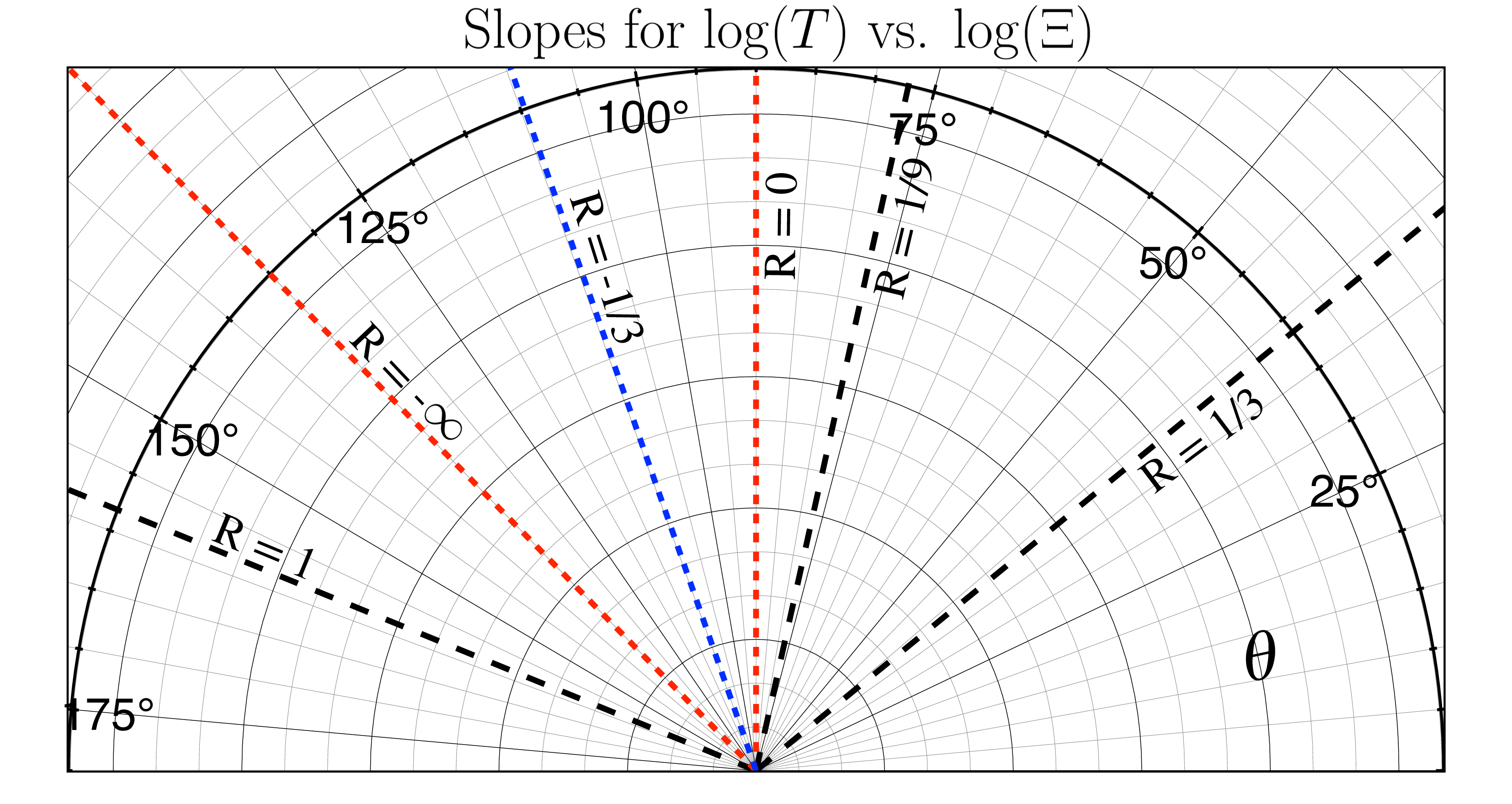}
\includegraphics[width=0.5\textwidth]{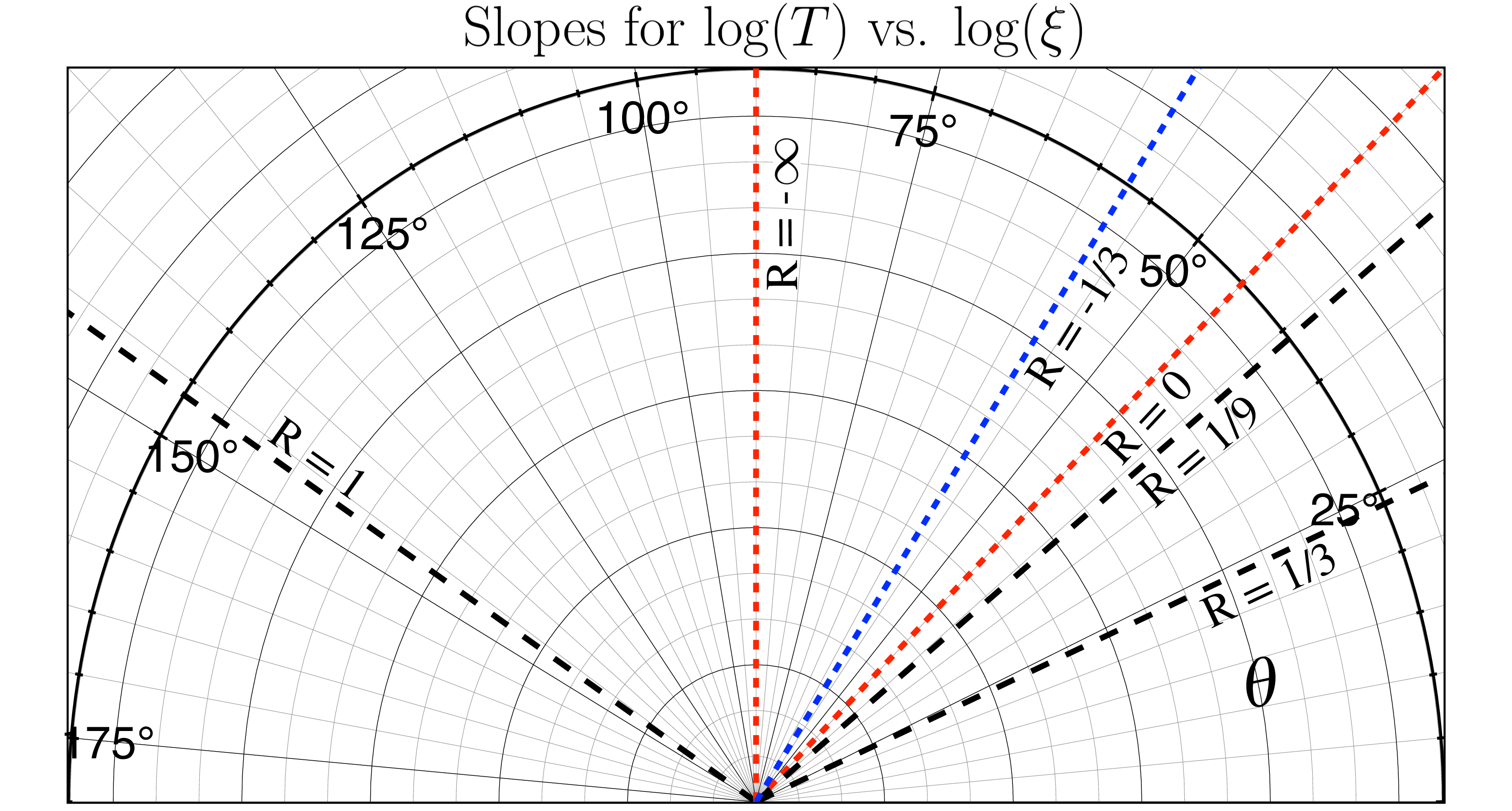}
\caption{Diagrams revealing the mapping between the critical $R$-values in Table~1 and the slopes on the $\log(T)$ vs. $\log(\Xi)$ (upper plot) and the $\log(T)$ vs. $\log(\xi)$ (lower plot) planes.  The dashed black lines mark the three $R$ values $1/9,1/3$, and $1$, while the red dashed lines mark $R=-\infty$ and $R =0$.  The slope of the dashed blue line is for our fiducial numerical example, that of the blue dot on our S-curves.}
\label{fig:slopes}
\end{figure}

\subsection{The slopes of S-curves}
We conclude our linear theory analysis by showing that the slopes of S-curves on the $[\log(T),\log(\Xi)]$ and $[\log(T),\log(\xi)]$-plane are just simple expressions involving the ratio $R$, so to some extent the above properties of TI can be ascertained just by analyzing S-curves.  
From standard thermodynamic relations (see again the appendix of Balbus 1995) and equation (B32), we find
\beq 
\begin{split}
\left[\frac{\partial \log(T)}{\partial \log(\Xi)}\right]_{\mathcal{L}} 
&= \f{N_\rho}{N_p} - 1 \\
&= \f{1}{\gamma R} - 1.
\end{split}
\label{Xi_slope}
\seq
In these derivations we set $\lambda_F = 0$ since conduction is not important for determining the stability of perturbations with $\lambda >> \lambda_F$.  The left hand side is the slope of any given point (which has $\mathcal{L} = 0$) on the S-curve.  
Noting that $d \log(\Xi) = d \log(\xi) - d \log(T)$, slopes in the $[\log(T),\log(\xi)]$-plane are given by
\beq 
\begin{split}
\left[\frac{\partial \log(T)}{\partial \log(\xi)}\right]_{\mathcal{L}} 
&= 1 - \f{N_p}{N_\rho}  \\
&= 1 - \gamma R.
\end{split}
\label{xi_slope}
\seq
In Figure~\ref{fig:slopes}, we plot slopes for all of the critical values of $R$ in Table~1 (dashed black lines).  These plots can be used to judge the nature of any local thermal instabilities along the S-curves.  For example, the blue dot in Figure~1 has $R \approx -1/3$, corresponding to the slope plotted as blue dashed lines in Figure~\ref{fig:slopes}.  It occupies the region in between the red dashed lines, which is the most common case --- the top row in Table~1.  We therefore know what to expect in the non-isobaric regime for these heating and cooling functions: all sound waves in the system will have $\lambda < \lambda_c$ and the entropy mode is always unstable.  

We note that the entire range $-\infty < R < 0$ occupies regions between the dashed red lines in these plots.  
It is instructive to consider a couple examples involving smooth variations of $R$ through a change in sign.  
Let us assume the plasma starts off either isobarically or isochorically unstable but not both, i.e. somewhere in between the red dashed lines.  
Then a counter-clockwise variation from within the $R=-\infty$ red line into the neighboring region with $R > 0$ can occur two ways: (i) by the plasma becoming \emph{both} isobarically and isochorically unstable, which is described by the 2nd from last row of Table~1, with the fast isochoric condensation mode being the only unstable mode; or (ii) by the plasma becoming isentropically unstable (once $N_p > 0$ or $N_\rho > 0$) with both acoustic modes becoming overstable (see the bottom row in Table~1).  Similarly, a clockwise variation from within the $R=0$ red line into the region with $0 < R < 1/9$ can occur in two ways: the plasma can either become (i) thermally stable or (ii) doubly unstable, implying a move from the first or second row in Table~1 to the third row, in which case there can be three unstable condensation modes in the system for some range of wavelengths.

\section{Simulations}
We explore the transition from isobaric to non-isobaric TI by running local simulations following the same methods developed by PW15, but neglecting the radiation forces.  We use the Gudonov code \athena (Stone et al., in prep.), employing the HLLC flux, piecewise-parabolic reconstruction, and the 3rd order accurate RK3 time-integration scheme.  Conduction is solved more accurately than in \oldathena (Stone et al. 2008) using interface fluxes (c.f. Choi \& Stone 2012).  Note that in the input file to \athenapp, $\mathtt{kappa\_iso}$ is a constant \emph{diffusivity}, whereas we have assumed a constant \emph{conductivity}, adopting the Spitzer value appropriate for our background equilibrium state (see PW15), which gives $\lambda_F = 0.19 \lambda_{th}$.  We therefore modified \athenapp's conduction module appropriately.  

The profiles in equation \eqref{linear_profiles} serve as the initial conditions for our simulations, with $A = 10^{-2}$ and $n_R$ obtained by solving equation \eqref{n_general} for the entropy mode for any given $k$.  We apply periodic boundary conditions.  The physical units are set by the initial position on the S-curve, marked with a blue dot in Figure~1, which are the same as in PW15.  Our resolution is always 256 zones/$\lambda_{th}$, which is sufficient to fully resolve cloud interfaces in a steady state, in which the interface width is about $\lambda_F$ independent of cloud size.

\begin{figure}
\includegraphics[width=0.45\textwidth]{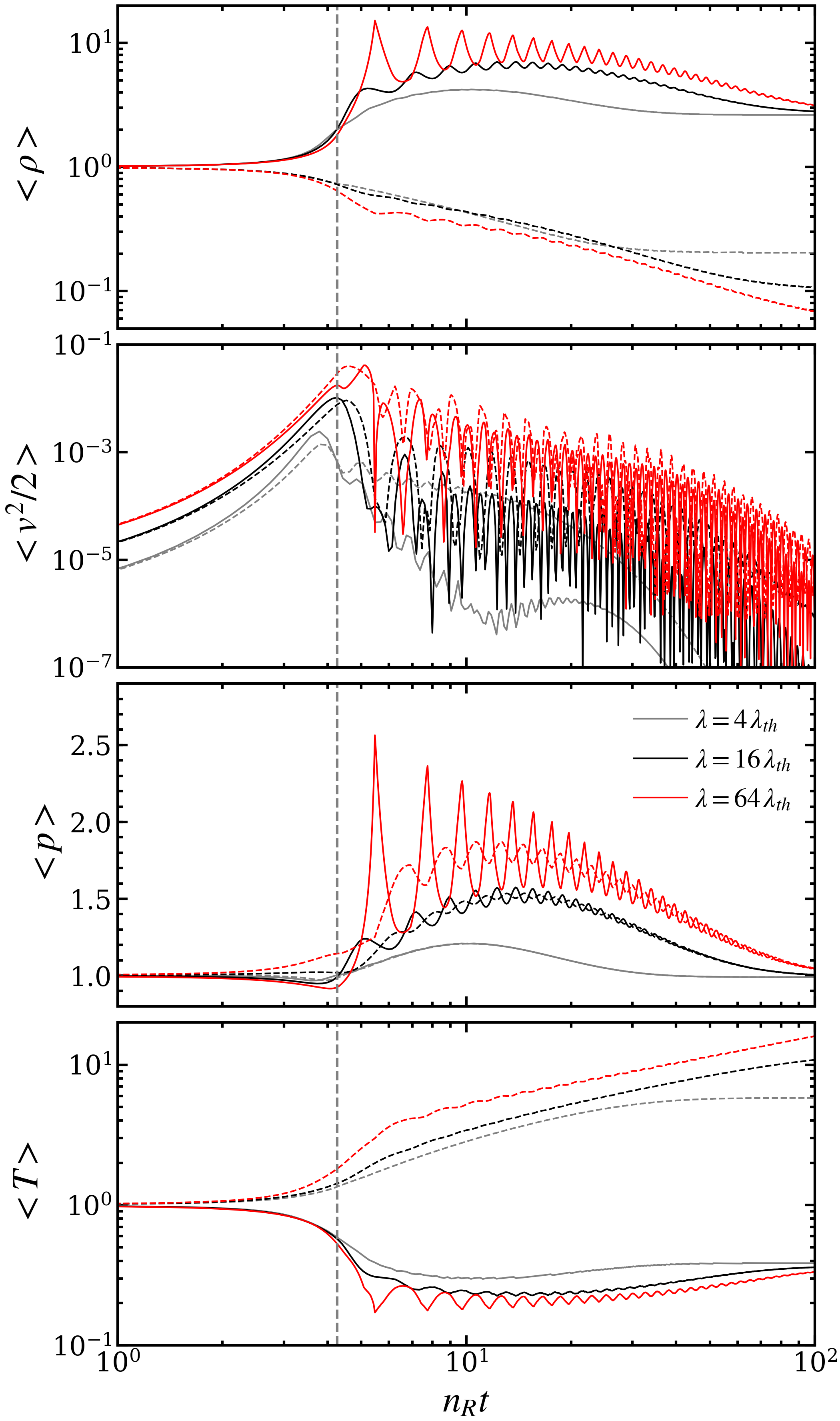}
\caption{Time evolution of spatially averaged quantities for clouds formed from three different size perturbations, $\lambda = 4\,\lambda_{th}$ (gray), $\lambda = 16\,\lambda_{th}$ (black), and $\lambda =  64\,\lambda_{th}$ (red). 
All quantities are in units of their background equilibrium values, except for $\langle v^2/2\rangle$, which is in units of $c_{s,0}^2$.
Solid and dashed lines indicate averages over the cooler and warmer phases, defined as $T <T_0$ and $T>T_0$, respectively.  
The gray vertical line marks the first appearance of saturation for the $\lambda = 16\,\lambda_{th}$ and $\lambda = 64\,\lambda_{th}$ clouds.
Notice we plot the $e$-folding time, $n_Rt$, not the actual time, to show that saturation occurs slightly later for the non-isobaric clouds.  Also, the non-isobaric clouds have smaller growth rates, $n_R$, so the periods of their oscillations and their damping timescales are $n_R^{-1}$ longer.
These $n_R^{-1}$ factors are $(4.75, 7.50, 20.6)\,t_{cool}$ for $\lambda = (4\,\lambda_{th}, 16\,\lambda_{th}, 64\,\lambda_{th})$.}
\end{figure}

\begin{figure*}
\includegraphics[width=\textwidth]{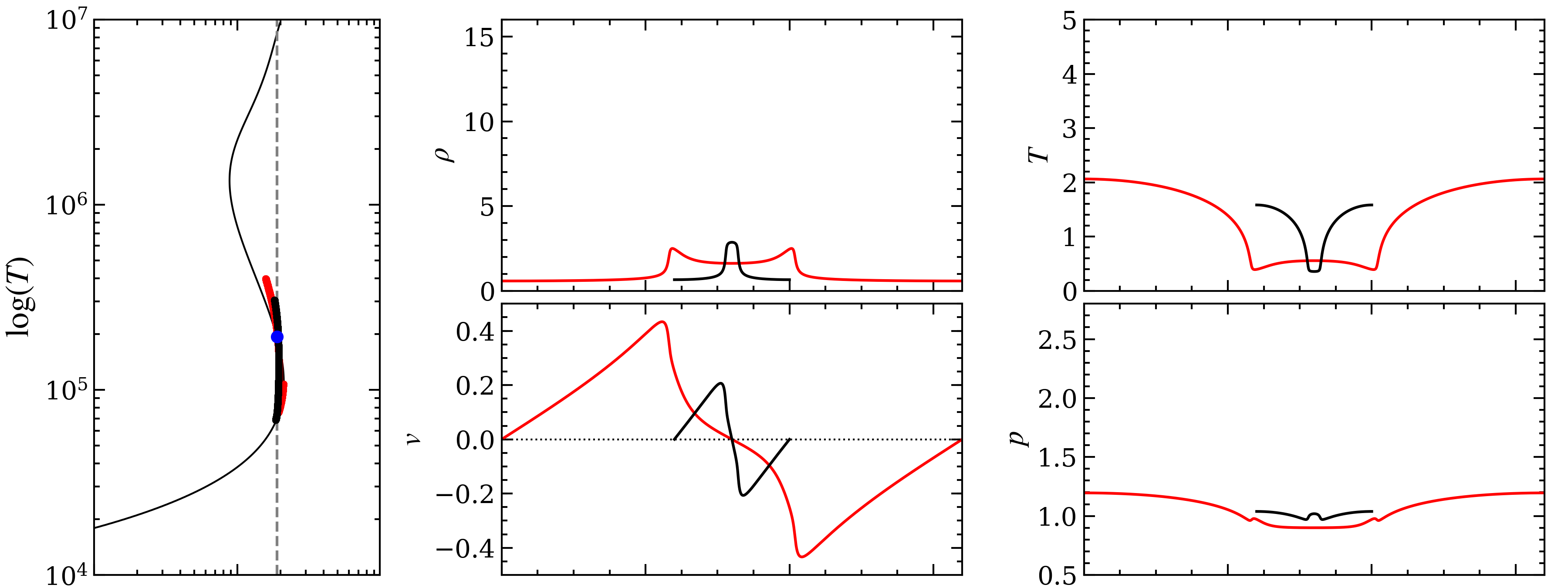}
\includegraphics[width=\textwidth]{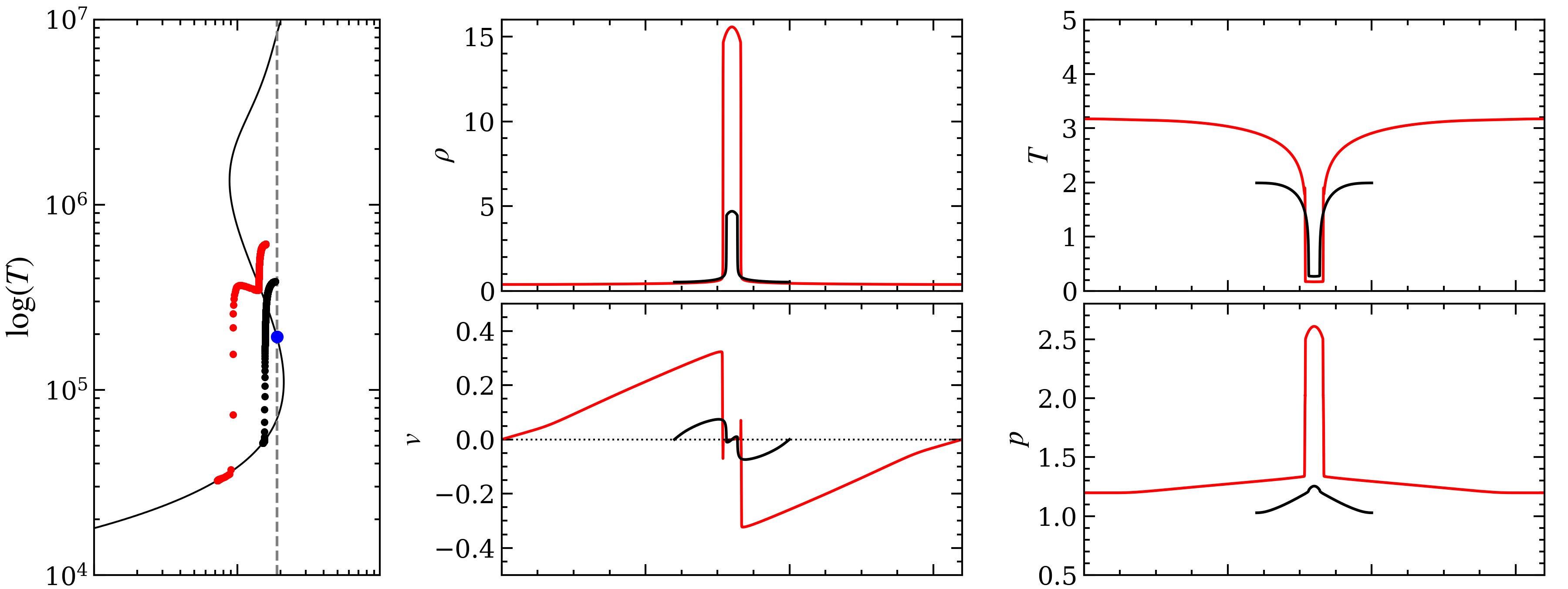}
\includegraphics[width=\textwidth]{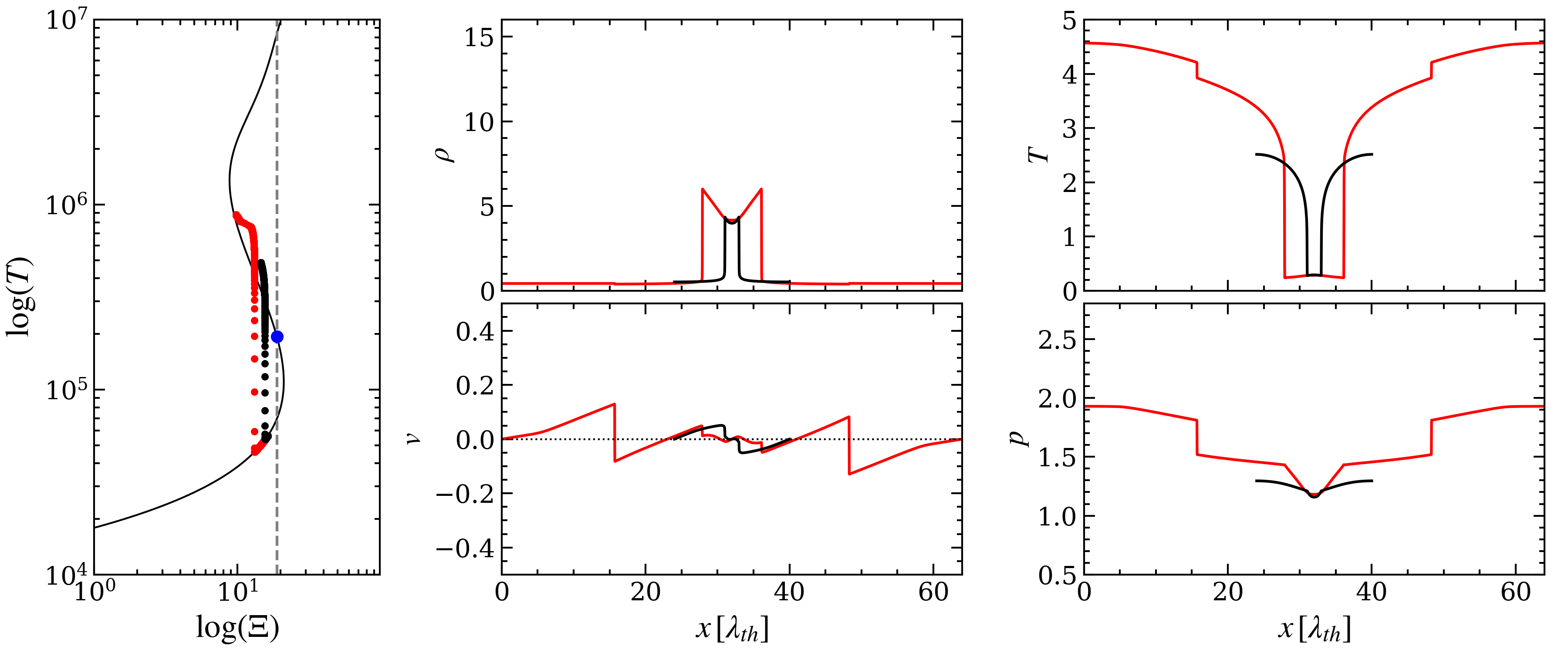}
\caption{Phase plots and spatial profiles of $(\rho, v, p, T)$ (in the same units used in Figure~3) during the early nonlinear phase of evolution for the $\lambda = 16\,\lambda_{th}$ and $\lambda = 64\,\lambda_{th}$ non-isobaric modes.  For clarity, we have centered the smaller cloud at $x = 32\,\lambda_{th}$.  Each panel corresponds to a single value of $nt$: the top set of panels correspond to the times marked with the gray vertical line in Figure~3, and the middle and bottom panels correspond to the first density peaks and troughs shown in Figure~3, respectively.
Notice that during the compression phases of an oscillation cycle (middle row), it becomes harder to resolve cloud interfaces.
The presence of large gaps in the points on the phase diagram are an indication that these interfaces are becoming under-resolved.    
The full simulation can be viewed at \url{http://www.physics.unlv.edu/astro/WP19aSims.html}.
}
\end{figure*}

 \begin{figure*}
\includegraphics[width=\textwidth]{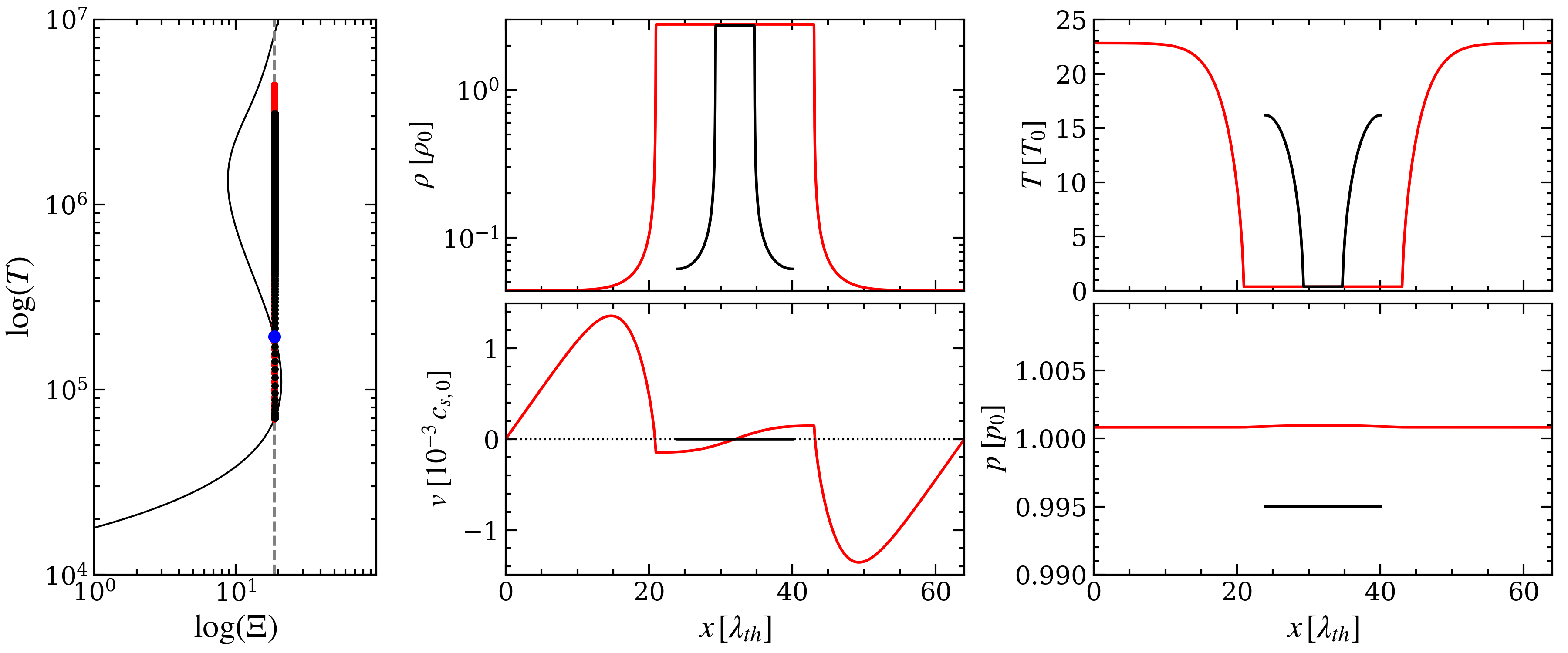}
\caption{As in Figure~4, but showing the nearly steady state profiles ($t = 2\times 10^3\,t_{cool}$) of the $\lambda = 16 \lambda_{th}$ and $\lambda = 64 \lambda_{th}$ non-isobaric modes.  The clouds have the same peak densities and minimum temperatures, corresponding to the same the cold-phase location on the S-curve (see the left panel).  
The velocity profile of the smaller cloud (black) is similar to the red profile but the velocity is much smaller, as larger clouds require larger advection velocities while still undergoing damped oscillations.}
\end{figure*}

Results for increasingly larger single mode perturbations are summarized in Figure~3, which shows the time evolution of the density, specific kinetic energy,  pressure, and temperature averaged over the hot and cold gas phases.  We compare a nearly isobaric cloud (gray, with $\lambda=4\,\lambda_{th}$) with two non-isobaric clouds, one (red, with $\lambda=64\,\lambda_{th}$) 4x larger than the other (black, with $\lambda=16\,\lambda_{th}$).  The overall evolution of non-isobaric perturbations is more complicated than isobaric ones but still very simple: the larger condensation modes both saturate into clouds that then undergo damped oscillations in their `struggle' to reach pressure equilibrium.  These oscillations excite sound waves (with $\lambda < \lambda_c$) as well as weak shocks (see the lower panels in Figure~4), which then dissipate their energy in the hot phase, possibly explaining why the temperature of the hot phase is higher for the $\lambda = 16\,\lambda_{th}$ mode and higher still for the $\lambda = 64\,\lambda_{th}$ mode.  

Notice that the oscillation periods and damping timescales of different non-isobaric clouds are essentially the same when plotted as a function of $e$-folding number.  This means that the damping rate and oscillation frequency decreases with increasing cloud size in proportion to $n_R$, i.e. the linear theory growth rate characterizes the long-term \emph{nonlinear} dynamics of non-isobaric clouds.  This is a natural consequence of the quantity $n_R/k$ being the characteristic velocity at which the cloud condenses at the end of the linear regime.  Since this velocity is proportional to $n_R$, the characteristic response to a sudden halting of the cloud growth is also proportional to $n_R$.   

To better understand the origin of the oscillations, we now take a close look at the initial nonlinear phase of TI for non-isobaric clouds.
The saturation of TI is a highly dynamical process during which the growing cloud core must suddenly halt its condensation and reach a new equilibrium at the cold phase.  As shown below, the gas in the core of the cloud does not all arrive at the cold phase at the same time, even for small clouds.  The halted growth must therefore be communicated across the entire cloud over some timescale that is proportional to $n_R^{-1}$.  In the isobaric regime, this timescale is on the order of the sound crossing time, and therefore sound waves can easily mediate the transition.  For large perturbations, however, the sound crossing time across the cloud exceeds the saturation timescale, meaning sound waves only communicate the changing dynamics throughout the cloud core gradually, and so oscillations ensue.

In Figure~4, we examine the spatial profiles during the early nonlinear phase for the quantities plotted in Figure~3.  In the top panels, we plot the very beginning of the saturation process, at times corresponding to the vertical gray line in Figure~3.  The left panel shows that some gas has reached the cold phase, while the density panel reveals where exactly this cold gas is located.  It is in the center of the cloud core for the $\lambda = 16\,\lambda_{th}$ cloud (there is a small peak in the black density profile at $x = 32$), while it is gas near the interfaces (the two red peaks) for the large cloud.  Returning to Figure~3, notice the vertical line coincides with a smaller initial peak in the specific kinetic energy panel (see the solid red line).  The maximum peak in $\langle v^2/2\rangle$ occurs later and corresponds to the remaining gas in the core reaching the stable cold phase.  
Comparing also with Figure~1, we see from the first velocity panel in Figure~4 that the peak condensation velocity is indeed $\approx n_R/k$.  

Later still, the first peaks in the mean density profiles in Figure~3 are reached.  This corresponds to times shown in the middle panels in Figure~4.  The intercloud gas continues to be further heated, as it is still far from the hot phase location on the S-curve.  (It need not reach the S-curve; evolution stops once $\kappa \nabla^2 T = \rho \mathcal{L}$, as shown by PW15.)  The pressure of the cloud is far out of equilibrium with the intercloud gas during these maximum compression cycles.  

The bottom set of panels in Figure~4 shows profiles for the first density troughs in Figure~3.  To conserve mass, the cloud width must increase during these expansion cycles, but this is accompanied by a strong compression of the intercloud gas, as shown by the higher temperature and pressure.  Higher pressure means lower $\Xi$, explaining the behavior of the `tracks' on the $[\log(T),\log(\Xi)]$-plane.  While clouds have no cohesive forces holding them together, the higher pressure of the intercloud gas supplies a restoring force to keep them in tact during this phase of the oscillation.  

These clouds eventually tend toward a steady state, as is evident from Figure~3.  This equilibrium state is shown in Figure~5.  Interestingly, despite the large difference in oscillation amplitudes that took place (e.g., pressure variations were more than 200\% for the larger cloud), both clouds settled back at the original pressure, corresponding to the same $\Xi$ as the blue dot on the S-curve.  Notice the final pressures deviate by less than 1\% from $p_0$.  The only indication that these clouds evolved differently is the higher intercloud gas temperature of the larger cloud.

\section{Discussion}
Our objective with this work was to determine how large, non-isobaric clouds evolve.  We ultimately aim to understand if such large clouds can survive longer than small isobaric clouds, and whether or not they are subject to a fragmentation process, such as the effect referred to as `shattering' in the work by M+18.  Here we discuss the relevance of our numerical simulations and our linear theory analysis 
to these questions.  We also compare our cloud sizes with the characteristic length scales discussed herein, as well as to the scale proposed to govern the size of individual shattered cloudlets.

\subsection{Shattering versus splattering}
In the previous section, we followed the evolution of non-isobaric individual entropy modes through the nonlinear regime.  What does it mean for a condensation mode to be `non-isobaric'?  
After all, by construction 
the solution remains practically isobaric, independent of wavelength, so long as $\delta p << p_0$.   
The classifier `non-isobaric' only takes on meaning in the nonlinear regime once there is a need to restore pressure balance.  

We emphasize this point because we have observed a numerical effect that can be easily mistaken for `shattering'.  
Namely, if we make the amplitude $A \equiv \delta \rho/\rho_0$ of the $\lambda = 64\,\lambda_{th}$ entropy mode perturbation $10^{-4}$ or smaller, we observed that many small scale (but still $>\lambda_F$) isobaric clouds formed and disrupted this large perturbation before it could reach the nonlinear regime.  
It appeared as though the $\lambda = 64\,\lambda_{th}$ mode spontaneously fragmented when really it was nothing more than unintended, fast growing isobaric perturbations overtaking this slow mode.  
Since this can occur in the linear regime when all modes evolve independently per the superposition principle, it cannot be the shattering process described by M+18.  
We verified that exponential growth of truncation errors of size $\sim 10^{-10}$ introduced by the finite precision of the initial growth rates or other input parameters is responsible for this occurrence.

This `numerical shattering' is most likely what is taking place in the simulations of M+18 (see their fig.~4).  
Our simulations have conclusively demonstrated that large clouds formed from the entropy mode are not subject to any sort of fragmentation, at least not in 1D when $R > -1$ (this exception is discussed in \S{5.2}).
As the perturbation wavelengths become increasingly larger than $\lambda_{th}$, the non-isobaric regime is entered gradually.  After the saturation of TI, sound waves become less and less able to communicate the condensation dynamics from the core of the cloud out to the interfaces.  The main dynamical consequence, as revealed in \S{4}, is the tendency for the cloud to undergo ever stronger oscillations in response to the sudden quenching of the cloud growth stage once the condensation reaches a stable cold phase in nearly thermal equilibrium.  In the absence of radiation forces (which we neglect here) or other additional processes, these oscillations then damp and a steady state mechanical equilibrium is approached.  

In consideration of this behavior, we now argue that there should be a fragmentation process operating in both isochorically unstable plasmas and isobarically unstable plasmas with $R \lesssim -1$ --- and one unrelated to M+18's shattering mechanism as they describe it.
Recall that the velocity of the condensing gas at the end of the linear regime is $n_R/k$, and this is bounded by $\sqrt{-R}\, c_{s,0}$ for isobarically unstable plasmas, while $n_R/k$ is unbounded for the fast isochoric mode since $n_R \rightarrow constant$ as $k \rightarrow 0$.  Thus, for a perturbation size large enough that its condensation velocity will exceed $c_{s,0}$ at the end of the linear regime, the resulting cloud could be fragmented during the first downward oscillation after the saturation phase of TI.  That is, the recoil pressure $\rho_c c_{s,0}^2$ will exceed the pressure of the confining gas (which, from the bottom panel of Figure~4, is only about $1.5 p_0$ during the initial saturation phase).  It seems appropriate to refer to this process as `splattering' because it is a dynamical response to a cloud's sudden `landing' on the cold phase of the S-curve.  Isochoric TI should lead to cloud splattering for any wavenumbers smaller than $k_{min} = n_R/c_{s,0} = (|N_\rho|/t_{cool})/c_{s,0}$, i.e. for any wavelengths larger than
\beq \lambda_{splatter} =  \f{2\pi}{|N_\rho|} \lambda_{th}, \label{lambda_splat} \seq
which is, to within a factor of a few, equal to $\lambda_c$.  
While we have not yet studied this process using numerical simulations, we speculate that it will lead to a breakup of the cloud into three pieces in 1D simulations: the core of the cloud will likely remain in tact as a portion of each side is ejected into the intercloud medium.

\subsection{Characteristic cloud sizes} 
This finding implies that there will be an upper limit to the sizes of condensations that undergo splattering.    
By mass conservation, a cloud that condensed out of a volume of gas $\sim \lambda^3$ with density $\rho_0$ must have a size
\beq d_c \sim \f{\lambda}{(\rho_c/\rho_0)^{1/D}},\label{d_c} \seq
where $\rho_c$ is the density at the stable cold phase and $D$ is the dimensionality of the system.  This formula would suggest that clouds can be as large as the largest perturbations in the system.  
Combining this with equation \eqref{lambda_splat}, we arrive at a characteristic size for clouds formed in isochorically unstable plasmas,
\beq (d_c)_{splatter} \sim  \f{2\pi}{|N_\rho|}\f{\lambda_{th}}{(\rho_c/\rho_0)^{1/3}} \text{\quad\quad (isochoric TI)} \label{d_c_max} .\seq
For condensations formed in isobarically unstable plasmas, 
there appears to be no upper limit to how large the clouds can be as far as local TI is concerned unless $R \lesssim -1$, in which case splattering can again occur for $k \lesssim 2\pi/\lambda_c$, thereby limiting cloud sizes to  
\beq (d_c)_{splatter} \sim \f{\lambda_c}{(\rho_c/\rho_0)^{1/3}} \text{\quad\quad (isobaric TI with $R \lesssim -1$)} \label{d_c_splatter} .\seq
For $R > -1$, a practical bound will be set by the constraints of the local approximation itself, e.g. the growth time $n_R^{-1}$ is required to be much less than the free fall time in a large scale accretion flow.  

We can compare $(d_c)_{splatter}$ to the characteristic scale of the individual cloudlets resulting from the shattering process of M+18:
\beq (d_c)_{shatter} \equiv \left(c_s\,t_{cool}\right)_{T_c}. \label{l_shatter}\seq
This scale is analogous to $\lambda_{th}$, but evaluated at $T_c$, the temperature of the cold phase on our S-curves, rather than $T_0$.  
In Figure~6, we compare this length scale with $(d_c)_{splatter}$ [from equation \eqref{d_c_splatter}] and $\lambda_F$ for our fiducial parameters.  The final cloud sizes measured from our simulations (black squares) for clouds formed from perturbations $\lambda = 2^m\lambda_{th}$ for $m=(0,1,\ldots,6)$ are also displayed, which are in excellent agreement with $d_c = \lambda/(\rho_c/\rho_0)$.  We also show $d_c$ evaluated with $D = 3$ (magenta stars) in equation \eqref{d_c} to illustrate how much larger clouds can become when simulated in 3D.  We note that these estimates were obtained without reference to simulation data, as we evaluated $\rho_c/\rho_0$ using our S-curve.  

\begin{figure}
\includegraphics[width=0.48\textwidth]{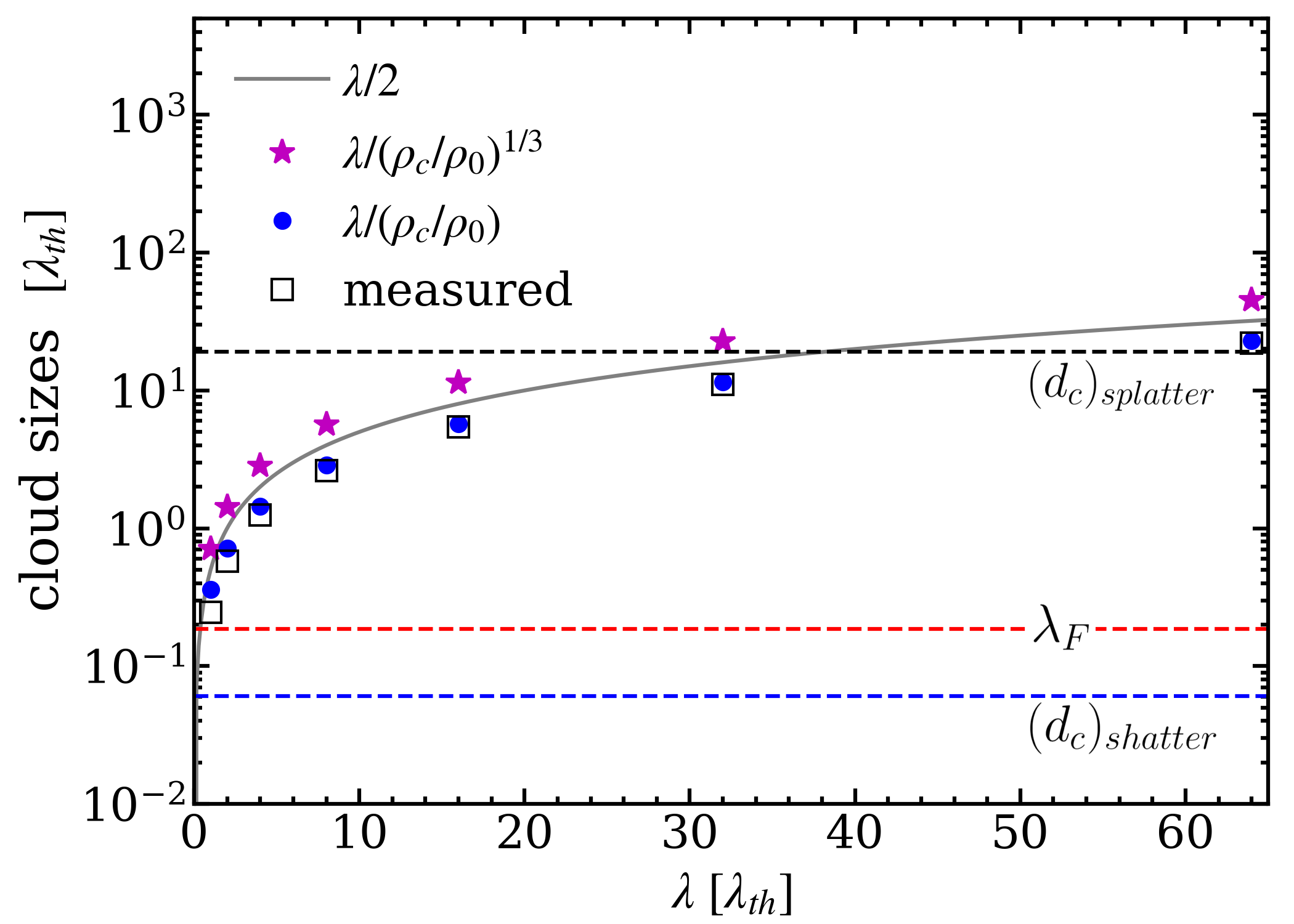}
\caption{Comparison of measured 1D steady state cloud sizes (squares) with predicted 1D (dots) and 3D (stars) sizes for clouds formed from perturbations with $\lambda = 2^m\lambda_{th}$ for $m=(0,1,\ldots,6)$.  For reference, the gray line shows $\lambda/2$, the length of the initial overdensity.  For our fiducial parameters, the upper dashed line marks $(d_c)_{splatter}$, the lower dashed line $(d_c)_{shatter}$, and the middle dashed line the Field length, $\lambda_F$.  Notice that $(d_c)_{shatter} << \lambda_F$.  
}
\end{figure}

Notice that there is a glaring issue with $(d_c)_{shatter}$ being a characteristic scale for shattered cloudlets: it is much less than the Field length and therefore such cloudlets would evaporate (Begelman \& McKee 1990; McKee \& Begelman 1991).
This is a general conclusion, not one peculiar to our fiducial simulation parameters, as plugging in the same values used below equation \eqref{thermal-Field-ratio} gives
\beq \f{(d_c)_{shatter}}{\lambda_F} = \f{17.5(\gamma - 1)^{-1}}{\eta (\rho_c/\rho_0)^{11/4}} \left(\f{T_c}{10^4\,\rm{K}}\right)^{-1/4} L_{23}^{-1/2}, \label{shatter-Field-ratio}\seq 
where $\eta \equiv L_c/L$ accounts for the different cooling rates for gas in the cold phase (cooling rate $L_c$) and gas undergoing evaporation (cooling rate $L$). 
For simplicity, in writing this ratio we assumed pressure equilibrium between the cold phase gas and the evaporating gas and that evaporation occurs at the temperature of the initial equilibrium state, $T_0$.  
Equation \eqref{shatter-Field-ratio} reveals that $(d_c)_{shatter}/\lambda_F$ is less than unity whenever the density contrast $\rho_c/\rho_0$ exceeds about $3.3$ for $\eta = 1$, $\gamma = 5/3$, $T_c = 10^4\,\rm{K}$, and $L =  10^{-23}\,\rm{erg\,cm^3\,s^{-1}}$.  Moreover, values of $\eta > 1$ will contribute to making $(d_c)_{shatter} < \lambda_F$. 

The recent simulations of wind-cloud interactions by Liang \& Remming (2018) are consistent with this finding, as their runs with thermal conduction showed that as their clouds were shredded by the Kelvin-Helmholtz instability, the cloud surface layers gradually evaporated.  While they concluded that their simulations were in support of the shattering hypothesis, there seems to be no clear evidence of cloud fragmentation, only surface disruption.

\subsection{Assessment of the various cases of TI} \label{sec:constraints}
We next discuss the applicability of all the different cases of TI listed in Table~1.  
The block in the first row with $\lambda < \lambda_c$ contains the most physically realizable unstable (`normal' herefafter) situation  --- isobaric instability of the entropy mode and two stable acoustic modes.  Our simulations fall into this category, and we focused on understanding the dynamics of entropy modes as $\lambda$ gradually increased beyond $\lambda_c$.  We found $\lambda_c$ to have no dynamical significance for the entropy mode (at least not when $R > -1$).  Rather, $\lambda_c$ merely dictates the maximum wavelength sound waves allowed within the system.  

All other categories in Table~1, except for the last row, involve isochoric instability.  Are all of these allowed to occur for realistic astrophysical heating and cooling functions?
A highly restrictive constraint is the following: collisionally ionized and photoionized plasmas will typically obey the criteria (e.g., Balbus \& Soker 1989),
\beq \pdLrho{T} > 0, \label{dLdrho_pos} \seq
for this is the statement that increasing the density at fixed temperature leads to increased cooling. 
We examine the implications of this constraint in detail in Appendix~B and simply quote the results of the analysis here.  
It turns out that for locations on the S-curve, this inequality is equivalent to $R > 1/\gamma$ when $N_\rho < 0$ and $R < 1/\gamma$ when $N_\rho > 0$, ruling out all but the first, fifth, and sixth rows of Table~1.  
That is, this inequality is automatically consistent with the `normal' instance of TI,  
in which an overdensity triggers runaway cooling leading to the growth of the entropy mode. 
However, the constraint that increased cooling must accompany compression is inconsistent with most instances of overstability in Table~1.  
Physically, what is needed for overstability is for the gas to be heated beyond its adiabatic rate over the course of a compression cycle, as this allows the amplitude to gradually increase.  

Thus, for any astrophysical net cooling functions satisfying inequality \eqref{dLdrho_pos}, we have arrived at the pleasingly simple result 
that the main possibility for \emph{isochoric} TI (namely, $R > 1/\gamma$) will be indicated by regions of negative slope on the $[\log(T),\log(\xi)]$-plane.
Since negative slopes on the $[\log(T),\log(\Xi)]$-plane have commonly been used to diagnose \emph{isobaric} instability in photoionized and collisionally ionized plasmas,
these two planes provide a complimentary means for a comprehensive stability analysis.  
However, it must also be explicitly checked that $N_\rho < 0$, as any positive value of $R$ can alternatively indicate a thermally stable region.\footnote{Scripts to compute to $N_\rho$ and $R$  for tabulated heating and cooling functions are available from the authors by request.} 

\subsection{Effects of magnetic fields}
While we did not take MHD into account, our linear theory results are still valid for perturbations of the entropy mode along field lines (Field 1965).  
Isobaric paths are no longer vertical lines on the $[\log(T),\log(\Xi)]$-plane in the presence of magnetic fields.  Rather, they are paths with slope $-1 - \beta/2$ (Bottorff et al. 2000), where $\beta$ is the ratio of gas pressure to magnetic pressure.  For example, the slope of our S-curve in the leftmost panel of Figure~1 is about -2.8 at the blue dot, whereas an isobaric path for a moderately strong magnetic field with $\beta = 1$ has a slope of $-1.5$.  There is a qualitative argument revealing that such a magnetic field is stabilizing for perturbations applied on the S-curve (e.g., Emmering et al. 1992; Bottorff et al. 2000): displacements confined to a path with a slope less steep than that of the S-curve (as in our example) are stable because a small increase in temperature puts the gas into a region of net cooling.  However, $N_p > 0$ is equivalent to Balbus' generalized isobaric stability criterion,
\beq \pdL{p} > \f{\mathcal{L}}{T}\seq
(when neglecting thermal conduction), which reveals that the above qualitative argument does not necessarily hold off the S-curve.  For displacements along this path but starting from a position above the S-curve (where $\mathcal{L} > 0$), the entropy mode is again prone to being unstable even though Field's criterion for stability is satisfied.  

In future work, we intend to extend our analysis of non-isobaric TI to fully account for the stability of MHD modes.  In the nonlinear regime, we expect that the tendency for non-isobaric clouds to undergo strong oscillations will lead to increased filamentary structure because magnetic tension will oppose oscillations perpendicular to the field. 
Additionally, the magnetic field will reduce the degrees of freedom for gas to condensate, so even fully 3D simulations of TI can result in characteristic cloud sizes given by equation \eqref{d_c} with $D < 3$.

\section{Summary and Conclusions}
Our linear theory analysis has identified all possible regimes of TI that follow from the dispersion relation found by Field (1965).
In our notation, the generalized isobaric and isochoric \emph{instability} criterions are $N_p < 0$ and $N_\rho < 0$, respectively, where $N_p$ and $N_\rho$ are defined in equation \eqref{NpNrho}.  
The following is a summary of results specific to the non-isobaric regime.

\begin{itemize}

\item The ratio of cooling rate derivatives, $R \equiv N_p/\gamma N_\rho$, is the basic parameter governing the different regimes of TI.  
Physically, $R$ is the ratio of the growth/damping rates of the isobaric ($n_R = -N_p t_{cool}^{-1}/ \gamma$) and fast isochoric ($n_R = -N_\rho t_{cool}^{-1}$) condensation modes.  
In the context of acoustic modes, $R\, c_{s,0}^2$ is the squared effective sound speed on the S-curve, $(\partial P/\partial \rho)_{\mathcal{L} = 0}$.

\item The only sound waves propagating within a thermally unstable plasma with $R < 0$ are those with $\lambda < \lambda_c$.
The acoustic modes become condensation modes for $\lambda > \lambda_c$.  Their stability is governed by the isochoric criterion and they have different condensation velocities ($n_R/k$), so we labeled these the fast and slow isochoric modes. 

\item  
The defining property of the entropy mode has to be altered in a plasma that is both isochorically and isobarically unstable, as it is no longer a non-propagating mode under most circumstances with $R > 0$, as shown in Table~1.  Rather, what defines the entropy mode is its stability criteria; it is governed by the isobaric criterion even in the limit $\lambda \rightarrow \infty$.  Likewise, the fast and slow isochoric modes always obey the isochoric criterion.  Any of these modes can become condensating or acoustic depending on the values of $R$ and $\lambda$.  

\item The slopes of S-curves on either the $[\log(T),\log(\xi)]$- or $[\log(T),\log(\Xi)]$-planes, while commonly used to diagnose isobaric instability, can only \emph{conclusively} diagnose instances of combined isobaric and isochoric instability with $0 < R < 1$ once $N_\rho$ is shown to be negative, since these same slopes could also indicate thermal stability.  They can then be used to \emph{distinguish} between the various TI regimes with $R>0$ (see Figure~2).

\item The constraint $(\partial\mathcal{L}/\partial\rho)_T > 0$ eliminates most possibilities of TI summarized in Table~1, leaving only rows 1, 5 and 6, as it corresponds to $R > 1/\gamma$ when $N_\rho < 0$ and $R < 1/\gamma$ when $N_\rho > 0$.  The two cases involving isochoric instability will appear as negative slopes on the $[\log(T),\log(\xi)]$-plane, just as isobaric instability shows up as negative slopes on the $[\log(T),\log(\Xi)]$-plane. 

\end{itemize}

We should emphasize that our linear theory analysis has fully accounted for the stability of regions \emph{off} the S-curves, through the appearance of $\mathcal{L}$ in the definition of $N_p$ [see equation \eqref{NpNrho}].  
This is important for several reasons.  First, as large clouds evolve in a multi-dimensional system, perturbations will naturally be generated with significant displacements off the S-curve (recall Figure~4).  Second, as explained in Appendix~B, 
regions of net heating may allow for outflows to become susceptible to the cases of double instability in Table~1 that are forbidden by the constraint $(\partial \mathcal{L}/\partial \rho)_T > 0$ for locations on the S-curve.  
Finally, the stability of magnetized plasmas can be better assessed, as it is known that sufficiently strong magnetic fields stabilize the entropy mode (e.g., Emmering et al. 1992), but only for perturbations applied on the S-curve (see \S{5.4}).

Since $R$ is the ratio of growth rates of the isobaric and fast isochoric mode (see the 1st bullet above), in the range $0 < R < 1/3$ any large perturbations present  will grow faster than small isobaric perturbations, leading to the formation of large clouds.  
By the last bulleted result, despite the plasma also being formally unstable by the isobaric criterion, there are actually only overstable acoustic modes for $1/\gamma< R < 1$ that can potentially compete with the fast isochoric modes (see row 5 of Table~1).  For $R > 1$, the acoustic modes are overdamped at all wavelengths and therefore it may be very difficult for pressure equilibrium to ever be established.  While the nonlinear outcome is unclear, such environments are likely to host only large non-isobaric clouds.  

Given that isobaric instability is far more common than isochoric instability, our simulations were aimed at understanding the dynamics of non-isobaric clouds formed from the entropy mode.  Despite ever stronger oscillations taking place as $\lambda$ increases, a single entropy mode perturbation always results in the formation of a single cloud.  
While the oscillations are a purely nonlinear effect, their frequencies are set by the \emph{linear} growth rates, as they are a response to the gas suddenly reaching the stable cold branch of the S-curve with a condensation velocity $n_R/k$.  

We showed that an interesting new phenomenon should accompany isochorically unstable plasmas and isobarically unstable plasmas with $R \lesssim -1$ --- cloud `splattering'.  This is a fragmentation process whereby the recoil triggered during the saturation phase of TI will not only lead to oscillations, but also to a break up of the cloud into several smaller clouds.  It is a mechanism that can potentially produce supersonic clumps starting from gas that is stationary in the comoving frame.  
We pointed out that the length scale of M+18's proposed shattered cloudlets will generally be less than the Field length, while the characteristic size of `condensation splatter' is expected to be $\approx \lambda_c$ [see equation \eqref{lambda_crit} and \S{5.2}].  
However, we should emphasize that M+18's cloudlets may not be subject to immediate evaporation in weakly collisional or collisionless plasma environments (e.g., in the CGM or the ICM), where thermal conduction can be significantly suppressed compared to the Spitzer value (e.g., Roberg-Clark et al. 2018; Komarov et al. 2018).

We conclude this paper with a mention of an interesting observational signature of non-isobaric multiphase gas dynamics.  
The largest clouds in our simulations oscillate wildly  
--- in size, density, and temperature --- until they dissipate their energy into the surrounding plasma.
If this damping phenomenon occurs in nature, perhaps within a complex of other large clouds of varying sizes, then each oscillating cloud would appear brightest when smallest (at the compression peaks), and the brightness peaks would be immediately followed by net blueshifts and preceded by net redshifts accompanying the expansion and compression phases, respectively. 
These oscillations occur at characteristic frequencies that decrease with increasing cloud size, implying that the corresponding timescales of different oscillating clouds within the complex would be on the order of the cloud sound crossing time.
This dynamics must therefore be inferred statistically in large-scale environments with long dynamical times --- ones that can be spatially resolved.  Integral field unit (IFU) observations would be particularly useful, as the marked increases and decreases in brightness over the course of an oscillation would be correlated with the relative blueshifts and redshifts that track the compression and expansion velocities.  No such correlations are expected for a turbulent multiphase environment, such as the ISM (and perhaps even the CGM; see Buie et al. 2018).  Computations of synthetic IFU observations for this signature of non-isobaric dynamics will be a focus of our future work on this topic.

\acknowledgments
TW thanks Hui Li, Jarrett Johnson, Patrick Kilian, and Randall Dannen for useful discussions related to this project, as well as
the organizers of the conference \emph{Multiphase AGN Feeding \& Feedback} (\url{http://www.sexten-cfa.eu/event/multiphase-agn-feeding-feedback/})
held in Sesto, Italy, where work on this project began.  
In particular, we thank Max Gaspari for feedback on the manuscript.  
Support for Program number HST-AR-14579.001-A was provided by NASA through a grant from the Space Telescope Science Institute, which is operated by the Association of Universities for Research in Astronomy, Incorporated, under NASA contract NAS5-26555.
TW is partially supported by 
the LANL LDRD Exploratory Research Grant 20170317ER.

\appendix

\section{Linearized solutions to the equations of non-adiabatic hydrodynamics}\label{sec:Analytic_solns}
Here we linearize the equations of gas dynamics to derive the dispersion relation in equation \eqref{DR},
as well as the analytic solutions quoted in equation \eqref{linear_profiles}.  
It is convenient to begin with the equations of hydrodynamics in the following form,
\beq
\pd{\rho}{t} + \mathbf{\nabla} \cdot \left(\rho \gv{v} \right) = 0 \label{eq:mass} .
\seq
\beq
\rho\left(\pd{\gv{v}}{t} + \gv{v}\cdot\grad{\gv{v}}\right) =  -\grad{p},
\label{eq:Euler}
\seq
\beq  \pd{\rho \mathcal{E}}{t} + \div{\rho \gv{v}  \mathcal{E}}= -p\div{\gv{v}} - \rho\mathcal{L} - \div{\H}. \label{eq:1stlaw_hydro} \seq
In the comoving frame of the plasma, the background velocity is $\gv{v}=0$, so the only non-vanishing velocity terms are those involving $\delta\gv{v}$ upon applying the Eulerian perturbation operator, $\delta$, to these equations.  
In the main text, we referred to background equlibrium quantities using a subscript `0', but there is no need for this notation here since perturbations are distinguished using $\delta$.  Upon eliminating $\mathcal{E}$ using the equation of state $\rho \mathcal{E} = p/(\gamma - 1)$, we find
\beq
\pd{\delta\rho}{t} + \mathbf{\nabla} \cdot \left(\rho \delta\gv{v}  \right) = 0 \label{eq:delta_mass}
\seq
\beq
 \rho\pd{\delta\gv{v}}{t} + \grad{\delta p} = 0 \label{eq:delta_mom}
\seq
\beq
 \f{1}{\gamma-1}\pd{\delta p}{t} + \f{1}{\gamma-1} \mathbf{\nabla} \cdot
\left(p \delta \gv{v} \right)  +
p \div{\delta\gv{v}} 
 = - \mathcal{L}\delta\rho - \rho\delta\mathcal{L} - \div{\delta \gv{q} }.
 \label{eq:delta_energy}
\seq
Since $\gv{q} = -\kappa(T) \grad{T}$, assuming Spitzer conductivity $\kappa(T) = \chi(T) T^{5/2}$ (Spitzer 1962), we formally have
\beq \delta \gv{q} = -\kappa(T)\left(\f{5}{2}\f{1}{T} +\f{1}{\chi}\f{d\chi}{dT} \right)\delta T\grad{T} - \kappa(T)\grad{\delta T} .\label{eq:delta_q}\seq
Note that the first term (with contributions from both the $T^{5/2}$-dependence and the $T$-dependence of the Coulomb logarithm) does not enter linear theory when perturbing a background flow with a uniform temperature.  We can take $\grad{T} = 0$ even if the background flow is dynamic so long as the temperature scale height $T/|\grad{T}|$ is much larger than the wavelength of the perturbation.  Similar considerations lead to 
$\mathbf{\nabla} \cdot \left(\rho \delta\gv{v}  \right) = \rho \mathbf{\nabla} \cdot \delta\gv{v}$ and
$\mathbf{\nabla} \cdot \left(p\delta\gv{v}  \right) = p \mathbf{\nabla} \cdot \delta\gv{v}$ in equations \eqref{eq:delta_mass} and \eqref{eq:delta_energy}, respectively.  
This local approximation also lets us discard a term from $\div{\delta \gv{q} }$, leaving just
$ \div{\delta \gv{q}} = - \kappa(T)\nabla^2{\delta T} $.

To close this set of equations, we need to write $\delta T$, as well as the $\delta \mathcal{L}$ appearing in \eqref{eq:delta_energy}, in terms of $\delta p$ and $\delta \rho$.  The functional form of $\mathcal{L}$ is $\mathcal{L} = \mathcal{L}(\rho, T)$, so we formally have 
\beq \delta\mathcal{L} = \pd{\mathcal{L}}{\rho}\delta\rho + \pd{\mathcal{L}}{T}\delta T. \seq
We can eliminate $\delta T$ in favor of $\delta p$ and $\delta \rho$ by perturbing the ideal gas law:
\beq \f{\delta T}{T} = \f{\delta p}{p} - \f{\delta \rho}{\rho}. \label{eq:deltaT} \seq
Combining the previous two equations gives
\beq 
\delta\mathcal{L} = T \mathcal{L}_{T,\rho} \left[\f{\delta p}{p} - \f{\delta \rho}{\rho} \right] + \rho \mathcal{L}_{\rho,T} \f{\delta \rho}{\rho} ,
\label{eq:dL}
 \seq
where we have adopted the notation from Balbus \& Soker (1989),
\beq \Theta_{A,B} \equiv \pdAB{A}{B} \quad\quad\text{(B held constant)}. \seq  
Finally, substituting in the plane wave solutions $\delta \rho = A_\rho \exp(nt + i\gv{k}\cdot\gv{x})$, $\delta\gv{v} = \gv{A}_v \exp(n t + i\gv{k}\cdot\gv{x}) $, and $\delta p = A_p \exp(n t + i\gv{k}\cdot\gv{x})$ exchanges the derivatives for $n$ and $i\gv{k}$, leading to
\beq
 n A_\rho + i\gv{k}\cdot(\rho \gv{A}_v ) = 0, \label{eq:Aq_mass}
\seq
\beq
n \rho \gv{A}_v + i\gv{k} A_p = 0 ,\label{eq:Aq_mom}
\seq
\beq
 \f{1}{\gamma-1}n A_p + \f{\gamma}{\gamma-1} i p \gv{k} \cdot \gv{A}_v
 =  -A_\rho \mathcal{L} - \rho T \mathcal{L}_{T,\rho} \left[\f{A_p}{p} - \f{A_\rho}{\rho} \right] + \rho^2 \mathcal{L}_{\rho,T} \f{A_\rho}{\rho} - \kk \kappa\,T \left[ \f{A_p}{p} - \f{A_\rho}{\rho} \right], 
 \label{eq:Aq_energy}
\seq
where $k^2 = \gv{k}\cdot \gv{k}$ in the last term is a result of the $\nabla^2$ operator.  
Equations \eqref{eq:Aq_mass},  \eqref{eq:Aq_mom}, and  \eqref{eq:Aq_energy} are made dimensionless by multiplying by $t_{cool}/\rho$, $t_{cool}/(\rho c_s)$, and $t_{cool}/(\rho c_s^2)$, respectively.  Further defining $n' = n t_{cool}$, $\gv{k}' = \gv{k} \lambda_{th}$, and the dimensionless amplitudes $\Arho = A_\rho/\rho$, $\Ap = A_p/\rho c_s^2$, and $\Avv= \gv{A}_v/c_s$ gives
\beq
n'\Arho + i\kp \cdot \Avv = 0, \label{eq:dAq_mass}
\seq
\beq
 n' \Avv + i\kp\Ap= 0, \label{eq:dAq_mom}
\seq
\beq
 \f{1}{\gamma-1}n' \Ap + \f{1}{\gamma-1} i\kp \cdot \Avv
  =   -t_{cool}\f{\Lambda}{c_s^2}\left[\f{T}{\Lambda} \mathcal{L}_{T,\rho}  -\f{\mathcal{L}}{\Lambda} + \f{\kkp}{\lambda_{th}^2} \f{\kappa\,T}{\rho \Lambda} \right]\gamma \Ap + t_{cool}\f{\Lambda}{c_s^2}\left[ \f{T}{\Lambda} \mathcal{L}_{T,p} +  \f{\kkp}{\lambda_{th}^2} \f{\kappa\,T}{\rho \Lambda}\right] \Arho , 
 \label{eq:dAq_energy}
\seq
where we have made use of the following identity to arrive at the final bracketed term above,
\beq \mathcal{L}_{T,p} = \mathcal{L}_{T,\rho} - \f{\rho}{T} \mathcal{L}_{\rho,T} .\seq 
Notice how $\lambda_F$ enters in equation \eqref{eq:dAq_energy} through the introduction of $\Lambda$ as a normalization for the cooling rate derivatives.  We defined the Field length as $\lambda_F =2\pi \sqrt{\kappa\,T/\rho \Lambda}$, rather than simply $\sqrt{\kappa\,T/\rho \Lambda}$ as originally done by Begelman and McKee (1990), in order to write
\beq  \f{\kkp}{\lambda_{th}^2} \f{\kappa\,T}{\rho \Lambda} = \left( \f{\lambda_F}{\lambda} \right)^2. \seq      
Using $c_s^2 = \gamma(\gamma-1)\mathcal{E}$, we have that $c_s^2/\Lambda = \gamma(\gamma - 1) t_{cool}$, so also note that both $t_{cool}$ and $\lambda_{th}$ cancel in equation \eqref{eq:dAq_energy}.  This is the mathematical justification that large wavelength perturbations are `just as isobaric' as small ones as far as linear theory is concerned.    

Equation \eqref{eq:dAq_energy} shortens significantly if we define new variables for the quantities in brackets, hence our definitions of $N_\rho$ and $N_p$ in equation \eqref{NpNrho}. 
Multiplying equation \eqref{eq:dAq_energy} by $\gamma(\gamma - 1)$ brings us to our final form for the perturbed energy equation,
\beq
\gamma n' \Ap  + i\gamma \kp \cdot \Avv =   -\gamma N_\rho \Ap + N_p \Arho . 
 \label{eq:dAq_energy2}
\seq

\subsection{Dispersion relation}
Taking the dot product of equation \eqref{eq:dAq_mom} with $\kp$ and then using equation \eqref{eq:dAq_mass} to substitute in for $\kp\cdot\Avv$ gives
\beq \f{\Ap}{\Arho} = -\left(\f{n'}{k'} \right)^2.\seq
Similarly, eliminating $\kp\cdot\Avv$ from equation \eqref{eq:dAq_energy2} also leads to an expression for $\Ap/\Arho$,
\beq \f{\Ap}{\Arho} = \f{n' + N_p/\gamma}{n' + N_\rho} .\seq
By equating these, we arrive at the cubic dispersion relation,
\beq n'\left(\f{n'}{k'}\right)^2 + N_\rho \left(\f{n'}{k'}\right)^2 + n' + \f{N_p}{\gamma} = 0, \label{eq:DR} \seq
which is equation \eqref{DR} upon returning to dimensional units.

Using $n = n_R + in_I$, the real and imaginary parts of equation \eqref{DR} can be written as
\beq (n_R + N_\rho/t_{cool})u^2 - (3 n_R + N_\rho/t_{cool}) w^2 + n_R c_s^2 + \gamma^{-1} (N_p/t_{cool}) c_s^2 = 0 ; \label{real}\seq
\beq n_I[3u^2 - w^2 + 2 (u/k) N_\rho/t_{cool} + c_s^2] = 0, \label{imag}\seq
where we have defined $u \equiv n_R/k$ and $w\equiv n_I/k$.  When $n_I = 0$, equation \eqref{imag} is trivially satisfied and equation \eqref{real} reduces to equation \eqref{n_general}.  When $n_I \neq 0$, equation \eqref{nR_Rpos} follows by combining equations \eqref{real} and \eqref{imag}. 

\subsection{Analytic solutions}
Returning to our notation of denoting background values with a `0' subscript, the time-dependent solutions according to the plane wave ansatz is 
\beq
\begin{split}
\rho(\gv{x},t) &= \rho_0 + A \rho_0 \exp({n\,t + i \gv{k}\cdot\gv{x}}), \\
v(\gv{x},t) &= A_v \exp({n\,t + i \gv{k}\cdot\gv{x}}), \\
p(\gv{x},t) &=  p_0 + A_p \exp({n\,t + i \gv{k}\cdot\gv{x}}) ,
\end{split}
\label{linear_solutions}
\seq
where $A = \Arho$ and $A_v \equiv \gv{k}\cdot \gv{A}_v /k$.  
By equations \eqref{eq:dAq_mass} and \eqref{eq:dAq_mom} in dimensional form, $A_v = i(n/k)A$ and $A_p = -\rho_0 (n/k)^2 A$.  
Further substituting $n = n_R + i n_I$, the analytic solutions become
\beq
\begin{split}
\rho(\gv{x},t) &= \rho_0 + A \rho_0 e^{n_R t} \cos({n_I t +  \gv{k}\cdot\gv{x}})], \\
v(\gv{x},t) &= - A e^{n_R t} \left[\f{n_I}{k}  \cos({n_I t +  \gv{k}\cdot\gv{x}}) + \f{n_R}{k}  \sin({n_I t + \gv{k}\cdot\gv{x}}) \right], \\
p(\gv{x},t) &=  p_0 + A \rho_0 e^{n_R t} \left[\f{2n_R n_I}{k^2}  \sin({n_I t + \gv{k}\cdot\gv{x}}) - \f{n_R^2 - n_I^2}{k^2}  \cos({n_I t + \gv{k}\cdot\gv{x}})  \right],
\end{split}
\label{linear_solutions}
\seq
The condensation mode solution quoted in equation \eqref{linear_profiles} is the special case $n_I = 0$.  On the other hand, isentropic sound wave solutions follow by setting $n_R = 0$, in which case $(\rho, v, p)$ all vary in phase.  Solutions with $n_R \neq 0$ and $n_I \neq 0$ describe either damped or overstable acoustic modes, depending on the sign of $n_R$.

\section{The constraint $(\partial \mathcal{L}/\partial \rho)_T > 0$}

Here we formally show how the constraint $(\partial \mathcal{L}/\partial \rho)_T > 0$ limits the cases of TI to rows 1, 5, and 6 in Table~1, at least for locations on the S-curve.    
Using the identity,
\beq \f{\rho_0}{T_0}\pdLrho{T} = \pdL{\rho} -  \pdL{p}, \seq
we can rewrite this constraint in terms of $N_\rho$ and $R$ to find
\beq N_\rho(1 - \gamma R) > \f{\mathcal{L}}{\Lambda_0}. \label{last_bound}\seq
In deriving this using equations \eqref{NpNrho}, notice that the Field length cancelled out, meaning the inferences from this bound hold irrespective of the strength of thermal conduction.  Similar to our analysis of isentropic instability, there are two cases to consider: $N_\rho < 0$ and $N_\rho > 0$.  For the isochorically stable case, we have
\beq R < \f{1}{\gamma} - \f{\mathcal{L}/\Lambda_0}{\gamma N_\rho} \quad\quad \text{(for $N_\rho > 0$)}. \seq
The top row of Table~1 is an $R < 0$ regime, so this condition is always satisfied unless the dynamical state of the flow is above the S-curve (in a net cooling regime with $\mathcal{L} > 0$) far enough to make $\mathcal{L}/\Lambda_0 > N_\rho$.  

The constraint $(\partial \mathcal{L}/\partial \rho)_T > 0$ is that compression leads to cooling, so we would expect isentropic instability to be ruled out.  
Indeed, the bottom row of Table~1 is an $R > 1$ regime, while this constraint places an upper limit on $R$ that is equal to $3/5$ for locations on the S-curve when $\gamma = 5/3$. 
Only below the S-curve in regions with sufficient heating, namely for $\mathcal{L}/\Lambda_0 < -2N_\rho/3$ when $\gamma = 5/3$, can acoustic modes be overstable.  

For the isochorically unstable case, inequality \eqref{last_bound} places a lower bound on $R$ instead:
\beq R > \f{1}{\gamma} + \f{\mathcal{L}/\Lambda_0}{\gamma |N_\rho|} \quad\quad \text{(for $N_\rho < 0$)}. \seq
For locations on the S-curve, this bound is simply $R > 1/\gamma$, ruling out rows 2, 3, and 4 in Table~1 when $\gamma = 5/3$.  
It may seem surprising that isochoric instabilty with $R < 0$ (second row of Table~1) is ruled out.  There is another way to reach this conclusion, namely via the identity 
\beq \pdLrho{T} =- \pdL{\rho} \left(\frac{\partial T}{\partial \rho}\right)_\mathcal{L} .\seq
The right hand side can be shown to have the same sign as $N_\rho [\partial \log(T)/\partial \log(\xi)]_\mathcal{L}$.  Since $N_\rho < 0$, to satisfy $(\partial \mathcal{L}/\partial \rho)_T > 0$  it must be that $[\partial \log(T)/\partial \log(\xi)]_\mathcal{L} < 0$, i.e. that the slope of the S-curve in the $[\log(T),\log(\xi)]$-plane is negative.  However, by equation \eqref{xi_slope}, this is impossible unless $R > 0$.  

The instances of TI in rows 2, 3, and 4 of Table~1 can only occur sufficiently below the S-curve in regions of net heating.  For example, accessing the regime in the third row of Table~1, where there are three unstable condensation modes at some wavelengths, would require $\mathcal{L} \lesssim -0.8|N_\rho|$.  Thermally driven outflows can require the gas to occupy regions below the S-curve (see the Compton heated wind models recently computed by Dyda et al. 2017).  Thus, the doubly unstable regimes of Table~1 are not necessarily inaccessible to dynamical flows.  


\vspace{5mm}

\begin{thebibliography}{}

\bibitem[Audit \& Hennebelle(2005)]{2005A&A...433....1A} Audit, E., \& Hennebelle, P.\ 2005, \aap, 433, 1
\bibitem[Balbus \& McKee(1982)]{1982ApJ...252..529B} Balbus, S.~A., \& McKee, C.~F.\ 1982, \apj, 252, 529
\bibitem[Balbus(1986)]{1986ApJ...303L..79B} Balbus, S.~A.\ 1986, \apjl, 303, L79 
\bibitem[Balbus(1985)]{1985ApJ...291..518B} Balbus, S.~A.\ 1985, \apj, 291, 518 
\bibitem[Balbus \& Soker(1989)]{1989ApJ...341..611B} Balbus, S.~A., \& Soker, N.\ 1989, \apj, 341, 611
\bibitem[Balbus(1995)]{1995ASPC...80..328B} Balbus, S.~A.\ 1995, The Physics of the Interstellar Medium and Intergalactic Medium, 80, 328 
\bibitem[Barai et al.(2011)]{2011MNRAS.418..591B} Barai, P., Proga, D., \& Nagamine, K.\ 2011, \mnras, 418, 591
\bibitem[Barai et al.(2012)]{2012MNRAS.424..728B} Barai, P., Proga, D., \& Nagamine, K.\ 2012, \mnras, 424, 728 
\bibitem[Begelman \& McKee(1990)]{1990ApJ...358..375B} Begelman, M.~C., \& McKee, C.~F.\ 1990, \apj, 358, 375 
\bibitem[Binney et al.(2009)]{2009MNRAS.397.1804B} Binney, J., Nipoti, C., \& Fraternali, F.\ 2009, \mnras, 397, 1804
\bibitem[Blondin(1994)]{1994ApJ...435..756B} Blondin, J.~M.\ 1994, \apj, 435, 756 
\bibitem[Bottorff et al.(2000)]{2000ApJ...537..134B} Bottorff, M.~C., Korista, K.~T., \& Shlosman, I.\ 2000, \apj, 537, 134
\bibitem[Brandenburg et al.(2007)]{2007ApJ...654..945B} Brandenburg, A., Korpi, M.~J., \& Mee, A.~J.\ 2007, \apj, 654, 945 
\bibitem[Brandenburg \& Nordlund(2011)]{2011RPPh...74d6901B} Brandenburg, A., \& Nordlund, {\AA}.\ 2011, Reports on Progress in Physics, 74, 046901
\bibitem[Br{\"u}ggen \& Scannapieco(2016)]{2016ApJ...822...31B} Br{\"u}ggen, M., \& Scannapieco, E.\ 2016, \apj, 822, 31
\bibitem[Buie et al.(2018)]{2018ApJ...864..114B} Buie, E., II, Gray, W.~J., \& Scannapieco, E.\ 2018, \apj, 864, 114 
\bibitem[Burkert \& Lin(2000)]{2000ApJ...537..270B} Burkert, A., \& Lin, D.~N.~C.\ 2000, \apj, 537, 270 
\bibitem[Choi \& Stone(2012)]{2012ApJ...747...86C} Choi, E., \& Stone, J.~M.\ 2012, \apj, 747, 86
\bibitem[Cowie \& McKee(1977)]{1977ApJ...211..135C} Cowie, L.~L., \& McKee, C.~F.\ 1977, \apj, 211, 135 
\bibitem[Cox(2005)]{2005ARA&A..43..337C} Cox, D.~P.\ 2005, \araa, 43, 337
\bibitem[David et al.(2014)]{2014ApJ...792...94D} David, L.~P., Lim, J., Forman, W., et al.\ 2014, \apj, 792, 94
\bibitem[Defouw(1970)]{1970ApJ...160..659D} Defouw, R.~J.\ 1970, \apj, 160, 659
\bibitem[Dyda et al.(2017)]{2017MNRAS.467.4161D} Dyda, S., Dannen, R., Waters, T., \& Proga, D.\ 2017, \mnras, 467, 4161
\bibitem[Emmering et al.(1992)]{1992ApJ...385..460E} Emmering, R.~T., Blandford, R.~D., \& Shlosman, I.\ 1992, \apj, 385, 460
\bibitem[Field(1965)]{1965ApJ...142..531F} Field, G.~B.\ 1965, \apj, 142, 531 
\bibitem[Field et al.(1969)]{1969ApJ...155L.149F} Field, G.~B., Goldsmith, D.~W., \& Habing, H.~J.\ 1969, \apjl, 155, L149 
\bibitem[Fukue \& Kamaya(2007)]{2007ApJ...669..363F} Fukue, T., \& Kamaya, H.\ 2007, \apj, 669, 363
\bibitem[Gazol et al.(2005)]{2005ApJ...630..911G} Gazol, A., V{\'a}zquez-Semadeni, E., \& Kim, J.\ 2005, \apj, 630, 911
\bibitem[Gaspari et al.(2012)]{2012ApJ...746...94G} Gaspari, M., Ruszkowski, M., \& Sharma, P.\ 2012, \apj, 746, 94
\bibitem[Gaspari(2015)]{2015MNRAS.451L..60G} Gaspari, M.\ 2015, \mnras, 451, L60
\bibitem[Gaspari \& S{\c a}dowski(2017)]{2017ApJ...837..149G} Gaspari, M., \& S{\c a}dowski, A.\ 2017, \apj, 837, 149 
\bibitem[Gaspari et al.(2017)]{2017MNRAS.466..677G} Gaspari, M., Temi, P., \& Brighenti, F.\ 2017, \mnras, 466, 677
\bibitem[Gnedin \& Hollon(2012)]{2012ApJS..202...13G} Gnedin, N.~Y., \& Hollon, N.\ 2012, \apjs, 202, 13
\bibitem[Gomez-Pelaez \& Moreno-Insertis(2002)]{2002ApJ...569..766G} Gomez-Pelaez, A.~J., \& Moreno-Insertis, F.\ 2002, \apj, 569, 766 
\bibitem[Hennebelle \& Passot(2006)]{2006A&A...448.1083H} Hennebelle, P., \& Passot, T.\ 2006, \aap, 448, 1083 
\bibitem[Heyvaerts(1974)]{1974A&A....37...65H} Heyvaerts, J.\ 1974, \aap, 37, 65
\bibitem[Inoue \& Omukai(2015)]{2015ApJ...805...73I} Inoue, T., \& Omukai, K.\ 2015, \apj, 805, 73
\bibitem[Iwasaki \& Inutsuka(2014)]{2014ApJ...784..115I} Iwasaki, K., \& Inutsuka, S.-i.\ 2014, \apj, 784, 115
\bibitem[Ji et al.(2018)]{2018MNRAS.476..852J} Ji, S., Oh, S.~P., \& McCourt, M.\ 2018, \mnras, 476, 852 
\bibitem[Kallman \& McCray(1982)]{1982ApJS...50..263K} Kallman, T.~R., \& McCray, R.\ 1982, \apjs, 50, 263 
\bibitem[Kim \& Narayan(2003)]{2003ApJ...596..889K} Kim, W.-T., \& Narayan, R.\ 2003, \apj, 596, 889 
\bibitem[Kim et al.(2013)]{2013ApJ...776....1K} Kim, C.-G., Ostriker, E.~C., \& Kim, W.-T.\ 2013, \apj, 776, 1
\bibitem[Kim \& Ostriker(2015)]{2015ApJ...802...99K} Kim, C.-G., \& Ostriker, E.~C.\ 2015, \apj, 802, 99
\bibitem[Kritsuk \& Norman(2002)]{2002ApJ...569L.127K} Kritsuk, A.~G., \& Norman, M.~L.\ 2002, \apjl, 569, L127 
\bibitem[Komarov et al.(2018)]{2018JPlPh..84c9005K} Komarov, S., Schekochihin, A.~A., Churazov, E., \& Spitkovsky, A.\ 2018, Journal of Plasma Physics, 84, 905840305
\bibitem[Kritsuk et al.(2017)]{2017NJPh...19f5003K} Kritsuk, A.~G., Ustyugov, S.~D., \& Norman, M.~L.\ 2017, New Journal of Physics, 19, 065003 
\bibitem[Krolik et al.(1981)]{1981ApJ...249..422K} Krolik, J.~H., McKee, C.~F., \& Tarter, C.~B.\ 1981, \apj, 249, 422 
\bibitem[Krolik(1999)]{1999agnc.book.....K} Krolik, J.~H.\ 1999, Princeton, N.~J.~: Princeton University Press, c1999.
\bibitem[Koyama \& Inutsuka(2004)]{2004ApJ...602L..25K} Koyama, H., \& Inutsuka, S.-i.\ 2004, \apjl, 602, L25
\bibitem[Kurosawa \& Proga(2009)]{2009MNRAS.397.1791K} Kurosawa, R., \& Proga, D.\ 2009, \mnras, 397, 1791
\bibitem[Lepp et al.(1985)]{1985ApJ...288...58L} Lepp, S., McCray, R., Shull, J.~M., Woods, D.~T., \& Kallman, T.\ 1985, \apj, 288, 58
\bibitem[Li \& Bryan(2014)]{2014ApJ...789..153L} Li, Y., \& Bryan, G.~L.\ 2014, \apj, 789, 153
\bibitem[Liang \& Remming(2018)]{2018arXiv180610688L} Liang, C.~J., \& Remming, I.~S.\ 2018, arXiv:1806.10688
\bibitem[Malagoli et al.(1987)]{1987ApJ...319..632M} Malagoli, A., Rosner, R., \& Bodo, G.\ 1987, \apj, 319, 632 
\bibitem[McCourt et al.(2012)]{2012MNRAS.419.3319M} McCourt, M., Sharma, P., Quataert, E., \& Parrish, I.~J.\ 2012, \mnras, 419, 3319
\bibitem[McCourt et al.(2018)]{2018MNRAS.473.5407M} McCourt, M., Oh, S.~P., O'Leary, R., \& Madigan, A.-M.\ 2018, \mnras, 473, 5407 (M+18)
\bibitem[McKee \& Begelman(1990)]{1990ApJ...358..392M} McKee, C.~F., \& Begelman, M.~C.\ 1990, \apj, 358, 392
\bibitem[McNamara et al.(2014)]{2014ApJ...785...44M} McNamara, B.~R., Russell, H.~R., Nulsen, P.~E.~J., et al.\ 2014, \apj, 785, 44
\bibitem[Meece et al.(2015)]{2015ApJ...808...43M} Meece, G.~R., O'Shea, B.~W., \& Voit, G.~M.\ 2015, \apj, 808, 43
\bibitem[Mehdipour et al.(2016)]{2016A&A...596A..65M} Mehdipour, M., Kaastra, J.~S., \& Kallman, T.\ 2016, \aap, 596, A65
\bibitem[Mo{\'s}cibrodzka \& Proga(2013)]{2013ApJ...767..156M} Mo{\'s}cibrodzka, M., \& Proga, D.\ 2013, \apj, 767, 156
\bibitem[Nakayama \& Masai(2001)]{2001A&A...375..328N} Nakayama, M., \& Masai, K.\ 2001, \aap, 375, 328
\bibitem[Nekrasov(2011)]{2011ApJ...739...88N} Nekrasov, A.~K.\ 2011, \apj, 739, 88
\bibitem[Parker(1953)]{1953ApJ...117..431P} Parker, E.~N.\ 1953, \apj, 117, 431
\bibitem[Piontek \& Ostriker(2004)]{2004ApJ...601..905P} Piontek, R.~A., \& Ostriker, E.~C.\ 2004, \apj, 601, 905
\bibitem[Piontek \& Ostriker(2005)]{2005ApJ...629..849P} Piontek, R.~A., \& Ostriker, E.~C.\ 2005, \apj, 629, 849
\bibitem[Prasad et al.(2015)]{2015ApJ...811..108P} Prasad, D., Sharma, P., \& Babul, A.\ 2015, \apj, 811, 108
\bibitem[Proga \& Waters(2015)]{2015ApJ...804..137P} Proga, D., \& Waters, T.\ 2015, \apj, 804, 137
\bibitem[Pulido et al.(2018)]{2018ApJ...853..177P} Pulido, F.~A., McNamara, B.~R., Edge, A.~C., et al.\ 2018, \apj, 853, 177 
\bibitem[Roberg-Clark et al.(2018)]{2018PhRvL.120c5101R} Roberg-Clark, G.~T., Drake, J.~F., Reynolds, C.~S., \& Swisdak, M.\ 2018, Physical Review Letters, 120, 035101 
\bibitem[Russell et al.(2014)]{2014ApJ...784...78R} Russell, H.~R., McNamara, B.~R., Edge, A.~C., et al.\ 2014, \apj, 784, 78
\bibitem[Russell et al.(2016)]{2016MNRAS.458.3134R} Russell, H.~R., McNamara, B.~R., Fabian, A.~C., et al.\ 2016, \mnras, 458, 3134
\bibitem[Sabano \& Yoshii(1977)]{1977PASJ...29..207S} Sabano, Y., \& Yoshii, Y.\ 1977, \pasj, 29, 207
\bibitem[Shadmehri et al.(2010)]{2010Ap&SS.326...83S} Shadmehri, M., Nejad-Asghar, M., \& Khesali, A.\ 2010, \apss, 326, 83
\bibitem[Sharma et al.(2010)]{2010ApJ...720..652S} Sharma, P., Parrish, I.~J., \& Quataert, E.\ 2010, \apj, 720, 652
\bibitem[Sharma et al.(2012)]{2012MNRAS.420.3174S} Sharma, P., McCourt, M., Quataert, E., \& Parrish, I.~J.\ 2012, \mnras, 420, 3174
\bibitem[Spitzer(1962)]{1962pfig.book.....S} Spitzer, L.\ 1962, Physics of Fully Ionized Gases, New York: Interscience (2nd edition), 1962 
\bibitem[Stern et al.(2016)]{2016ApJ...830...87S} Stern, J., Hennawi, J.~F., Prochaska, J.~X., \& Werk, J.~K.\ 2016, \apj, 830, 87
\bibitem[Stiele et al.(2006)]{2006MNRAS.372..862S} Stiele, H., Lesch, H., \& Heitsch, F.\ 2006, \mnras, 372, 862 
\bibitem[Stocke et al.(2013)]{2013ApJ...763..148S} Stocke, J.~T., Keeney, B.~A., Danforth, C.~W., et al.\ 2013, \apj, 763, 148
\bibitem[Stone et al.(2008)]{2008ApJS..178..137S} Stone, J.~M., Gardiner, T.~A., Teuben, P., Hawley, J.~F., \& Simon, J.~B.\ 2008, \apjs, 178, 137
\bibitem[Sutherland \& Dopita(1993)]{1993ApJS...88..253S} Sutherland, R.~S., \& Dopita, M.~A.\ 1993, \apjs, 88, 253 
\bibitem[Temi et al.(2018)]{2018ApJ...858...17T} Temi, P., Amblard, A., Gitti, M., et al.\ 2018, \apj, 858, 17
\bibitem[Tremblay et al.(2016)]{2016Natur.534..218T} Tremblay, G.~R., Oonk, J.~B.~R., Combes, F., et al.\ 2016, \nat, 534, 218
\bibitem[Vantyghem et al.(2016)]{2016ApJ...832..148V} Vantyghem, A.~N., McNamara, B.~R., Russell, H.~R., et al.\ 2016, \apj, 832, 148
\bibitem[V{\'a}zquez-Semadeni et al.(2000)]{2000ApJ...540..271V} V{\'a}zquez-Semadeni, E., Gazol, A., \& Scalo, J.\ 2000, \apj, 540, 271 
\bibitem[Voit et al.(2015)]{2015Natur.519..203V} Voit, G.~M., Donahue, M., Bryan, G.~L., \& McDonald, M.\ 2015, \nat, 519, 203
\bibitem[Voit et al.(2017)]{2017ApJ...845...80V} Voit, G.~M., Meece, G., Li, Y., et al.\ 2017, \apj, 845, 80 
\bibitem[Wagh et al.(2014)]{2014MNRAS.439.2822W} Wagh, B., Sharma, P., \& McCourt, M.\ 2014, \mnras, 439, 2822 
\bibitem[Waters \& Proga(2016)]{2016MNRAS.460L..79W} Waters, T., \& Proga, D.\ 2016, \mnras, 460, L79 
\bibitem[Werk et al.(2013)]{2013ApJS..204...17W} Werk, J.~K., Prochaska, J.~X., Thom, C., et al.\ 2013, \apjs, 204, 17
\bibitem[Xia et al.(2014)]{2014ApJ...792L..38X} Xia, C., Keppens, R., Antolin, P., \& Porth, O.\ 2014, \apjl, 792, L38 
\bibitem[Xia \& Keppens(2016)]{2016ApJ...823...22X} Xia, C., \& Keppens, R.\ 2016, \apj, 823, 22
\bibitem[Yoneyama(1972)]{1972PASJ...24...87Y} Yoneyama, T.\ 1972, \pasj, 24, 87 
\bibitem[Zanstra(1955)]{1955VA......1..256Z} Zanstra, H.\ 1955, Vistas in Astronomy, 1, 256
\end{thebibliography}
\end{document}